\documentclass[review,fleqn]{elsarticle}
\usepackage{hyperref}
\pdfoutput=1
\usepackage{natbib}
\usepackage{amsthm}
\usepackage{amsmath}
\numberwithin{equation}{section}
\usepackage{subfig}
\usepackage{wrapfig}
\usepackage{setspace}
\usepackage{mathrsfs}
\usepackage[style=base]{caption}
\journal{Mathematical Biosciences}
\begin{document}
\begin{frontmatter}
\title{Random and non-random mating populations: Evolutionary dynamics in meiotic drive}
\author{Bijan Sarkar \corref{cor}}
\address{Department of Mathematics, Neotia Institute of Technology, Management and Science, Diamond Harbour Road, 24 Parganas (South), Pin-743368, West Bengal, India}
\cortext[cor]{E-mail: bijan0317@yahoo.com, bijan0317@gmail.com}

\begin{abstract}
Game theoretic tools are utilized to analyze a one-locus continuous selection model of sex-specific meiotic drive by considering nonequivalence of the viabilities of reciprocal heterozygotes that might be noticed at an imprinted locus. The model draws attention to the role of viability selections of different types to examine the stable nature of polymorphic equilibrium. A bridge between population genetics and evolutionary game theory has been built up by applying the concept of the Fundamental Theorem of Natural Selection. In addition to pointing out the influences of male and female segregation ratios on selection, configuration structure reveals some noted results, e.g., Hardy-Weinberg frequencies hold in replicator dynamics, occurrence of faster evolution at the maximized variance fitness,  existence of mixed Evolutionarily Stable Strategy ($ESS$) in asymmetric games, the tending evolution to follow not only a $1:1$ sex ratio but also a $1:1$ different alleles ratio at particular gene locus. Through construction of replicator dynamics in the group selection framework, our selection model introduces a redefining bases of game theory to incorporate non-random mating where a mating parameter associated with population structure is dependent on the social structure. Also, the model exposes the fact that the number of polymorphic equilibria will depend on the algebraic expression of population structure.  
\end{abstract}


\end{frontmatter}


\section{Introduction} 
Mendel demonstrates that a hybrid between two different varieties possesses both types of parental factors in the gametes (the principle of segregation) and the two individual genes in a particular gene pair ($2$ alleles) are equally represented in its gametes. Segregation distorters, however, violate this rule by biasing segregation in their favor. Generally, Segregation distortion refers to {\itshape {any distortion of meiosis or gametogenesis such that one of a pair chromosomes in a heterozygote is recovered in greater than half of the progeny}} \citep{hurst2001}. The associated gene is said to ``drive" and the phenomenon of meiotic drive is the subset of segregation distortion. In an extended sense, a segregation distorter/meiotic drive gene is nothing but the paradigmatic category of selfish genetic elements because such elements are either neutral or detrimental to the organism's fitness \citep{Burt2006}. The genes found in an inversion on chromosome $2$ of Drosophila melanogaster and $t$-complex in mice are tailored to explain the segregation distortion mechanism.

As segregation distorters occur in many species and are likely to be very common \citep{hurst2001,Jaenike1996}, a fascinating mathematical theory has been developed through mathematical models to answer the questions concerning the stability and evolution of Mendelian segregation (see Feldman and Otto \citep{Feldman1991}, Haig and Grafen \citep{Haig1991}, Weissing and van Boven \citep{Weissing2001}, \'{U}beda and Haig \citep{Ubeda2004} and references therein). \'{U}beda and Haig \cite{Ubeda2004}  were the first population geneticists to formulate a one-locus sex-specific segregation distortion model by including a genomic imprinting concept. They provided a numerical analysis of the equilibria, finding examples of parameter sets with three polymorphic equilibria in which at most two of them were stable. It was also noted that, in general, population mean fitness is not maximized at polymorphic equilibria.

It is seen that since the main stream of studies of evolutionary game dynamics basically concentrates on the evolution of strategies in animal conflicts, the evolutionary dynamics of population genetic mechanisms is often neglected in the context of evolutionary game theory. In the majority of cases, neglecting the underlying genetic architecture in detail, the models of evolutionary structure are developed in the mentioned research field where it has been tried to set up a relation between population genetics and evolutionary game theory through the perception of evolutionary stable strategies and quantitative genetics \citep{Hines1980,Hines87,Abrams1993}. Most of the evolutionary game theorists believe that, besides the evolution of phenotypes, game dynamics is only directly applicable to study of a single locus haploid genetic model; consequently, genetics related focus has mainly been confined in haploid structure \citep{MaynardSmith1982,Hammerstein1996,Hofbauer1998,Huang2010,Broom2013,Ohtsuki2014}. However, it is frequently argued that it can equally be applied to study of the more complex genetic evolution than asexual. In the recent years, the concepts of stochastic evolutionary dynamics \citep{Nowak2004,Traulsen2006,Altrock2010} and coalescence theory \citep{Antal2009,Nowak2010,Allen2012} have built a stable foundation between evolutionary game theory and population genetics.

A bridge between population genetics and game theory is constructed by a simple relabeling of terms. Any single-locus, diploid model in population genetics can be interpreted as a game in which individuals are the players and distinct alleles are the strategies \citep{Hofbauer1982,Cressman2003,Traulsen2012} where the frequencies of genes act as meanfitness optimizer \citep{Haig2014}. However, based on different framework structure, game theory for diploid population has also  been studied in many literatures \citep{Smith1987,Rowe1988,Lessard2005,Hashimoto2009}. Recently, in the context of meiotic derive Traulsen and Reed \cite{Traulsen2012} have formulated  a model of diploid population by considering an interaction between alleles in a diploid genome as a two player game. The dynamics is studied through the well-known replicator equation and meiotic derive has been explained as a social dilemma such as the prisoner's dilemma or the snowdrift game. Cyclic dominance found in the rock-paper-scissors game is also embedded in their model.

In general, the evolutionary game theorists use the replicator dynamics to considering game with $2$ player or many players in well-mixed large population because its equilibrium concept, an evolutionarily stable strategy ($ESS$) -- refinement of Nash equilibrium, $ESS$ is an asymptotically stable state of the evolutionary dynamics -- positively describes evolutionary outcomes in environments. However, there are many population games with non-random matching, concerning to games of group selection \citep{Bergstrom2002,Bergstrom2003,Wilson2007} where formation shows some assortment. At this point, we can not deny that the group selection is not the correct expression for what is just non-random matching because the non-random can be coped with in the usual evolutionary framework, in the way in which the concept of non-random matching merely needs to be built into the definition of the game (see Taylor and Nowak \cite{Taylor2006}). In a particular sense, matching is typically assortative meaning that individuals have a higher probability of being matched with other like-natured individuals than with different-natured individuals.

Since the population structure plays a crucial role in evolutionary theory, the canonical group selection models \citep{Smith1964, Wilson1975,Wilson1977} can also influence the genetic literature. The canonical group selection model is called either the haystack model or the multi-level selection model. In order to construct the relation between group selection replicator dynamics and population structure, van Veelen \cite{Veelen2011} partitions the whole population into groups of size $n$, within which $n$-player game is played where group state representing the group frequencies of the different types of the groups in the population, forms the population structure. In the model, fitness is assigned to individuals (individualist perspective; see Kerr and Godfrey-Smith \cite{Kerr2002} and references therein) rather than to the groups in such a way that replicator dynamics depends only on average payoffs of the individuals similar to the game selection model. However, such models can formally recast so that groups are fitness bearing. It was Kerr and Godfrey-Smith \cite{Kerr2002} who provided a detailed analysis of the multi-level selection model in the two strategy case. In the recent work of Jensen and Rigos \cite{Jensen2012}, the thought of matching rule is elaborated to any number of strategies by means of a rigorous formalism. Using the analytical thinking of structural conception of replicator dynamics, the model framework draws a concrete line between game and group selection theories.

In this article, I extend a traditional model for selection at an autosomal imprinted locus in a sex-differentiated population, by considering the sex-specific viability with assigning separate viabilities to reciprocal  heterozygotes. Allowing overlapping generations, \'{U}beda and Haig \cite{Ubeda2004}'s discrete genetic framework has been used to develop  continuous governing equations of evolutionary dynamics, namely replicator dynamics where, implementing the transmission ratio distortion rules at an imprinted locus, the concept of segregation distortion is incorporated. The group selection version of replicator dynamics is also introduced to examine the non-random influence. Here, my main intention is to formulate a set of link-results between evolutionary game theory and population genetics following the path constructed by Traulsen and Reed \cite{Traulsen2012}.

\section{The model of evolutionary dynamics} 

We pay attention to a specific type of evolutionary dynamics, called replicator dynamics to address the question, whether a polymorphic population with the two - population profiles can also be stable by generating the $ESS$. In the evolutionary game, it is assumed that individuals are programmed to use only pure strategies and passes this behavior to its descendants without modification; because of that the individuals in such selection dynamics are called replicators, existed in several different types \citep{Hofbauer1998,Broom2013,Taylor1978,Weibull1997}. As the success of evolutionary game dynamics is defined by the fitness of individuals, considering a pairwise contest population game with different action sets and different payoff matrices, in the context of well-mixed populations, the evolutionary game dynamics can be written as 
\begin{eqnarray}
\dot{x}_{i}&=&x_{i} ( \pi_{1}(e^{i},Y)-\pi_{1}(X,Y)),\nonumber \\
\dot{y}_{i}&=&y_{i} ( \pi_{2}(e^{i},X)-\pi_{2}(Y,X))\label{eq:1}
\end{eqnarray} 
where $\pi_{1}(X,Y))$ and $\pi_{2}(Y,X))$ signify the expected fitness payoffs of the two populations when the frequencies of types relating to strategy $e^{i}$ (unit vector) are $x_{i}$ and $y_{i}$, and corresponding average fitnesses of types are given by $\pi_{1}(e^{i},Y)$ and $\pi_{2}(e^{i},X)$.

Letting the pairwise contest population game being performed among alleles at a single locus, we move to the field of population genetics to establish that game theoretic perspective and population genetic perspective lead to exactly the same dynamics in respect of equations (\ref{eq:1})which govern allele frequencies. Consider an infinite, panmictic diploid population. The viability of genotypes $A_{1}A_{1}$, $A_{1}A_{2}$, $A_{2}A_{1}$ and $A_{2}A_{2}$ in males are designed as $W_{11}$, $W_{12}$, $W_{21}$ and $W_{22}$, with corresponding values $V_{11}$, $V_{12}$, $V_{21}$, $V_{22}$ in females where in males the autosomal allele written first has a paternal origin while the one written second has a maternal origin whereas paternally inherited alleles are listed second in the females. By explicitly considering perfect transmission of one allele in one sex and the other allele in the opposite sex, let the segregation ration of $A_{1}$ be $k$ in male meiosis and $\kappa$ in female meiosis, while the corresponding rations for $A_{2}$ are $(1-k)$ and $(1-\kappa)$ ($0<k, \kappa <1 $). That is values of $k$, $\kappa$ less than one-half can be interpreted as segregation distortion in favor of $A_{2}$ or negative segregation distortion of $A_{1}$ in respect of particular sex. The term ``drive" is used by Burt and Trivers \cite{Burt2006} to denote the greater than Mendelian (``super-Mendelian") transmission of a selfish genetic element, whereas ``drag" is the opposite implying less than Mendelian inheritance (``sub-Mendelian"). And hence if we let the frequency of gametes $A_1$, $A_2$ be $x_{t}$, $1-x_{t}$ in sperm and $y_{t}$, $1-y_{t}$ in eggs, then in the mating pool for the following generation the frequencies of male allele $A_1$s and female allele $A_{1}$s are \citep{Ubeda2004, Kidwell1977}
\footnotesize
\begin{eqnarray}
x_{t+\Delta t}&=& \frac{W_{11}x_{t}y_{t}+k(W_{12}x_{t}(1-y_{t})+W_{21}y_{t}(1-x_{t}))}{W_{11}x_{t}y_{t}+W_{12}x_{t}(1-y_{t})+W_{21}y_{t}(1-x_{t})+W_{22}(1-x_{t})(1-y_{t})},   \nonumber \\
y_{t+\Delta t}&=&\frac{V_{11}x_{t}y_{t}+\kappa (V_{12}x_{t}(1-y_{t})+V_{21}y_{t}(1-x_{t}))}{V_{11}x_{t}y_{t}+V_{12}x_{t}(1-y_{t})+V_{21}y_{t}(1-x_{t})+V_{22}(1-x_{t})(1-y_{t})}. 
\end{eqnarray}
\normalsize
Differential expression of genes depending on their parental origin is referred to genomic imprinting that can cause reciprocal heterozygotes to have distinguishable phenotypes and different viabilities. Here, considering $W_{12}$, $V_{21}$ being constants over time, we introduce the following  genomic imprinting relations:
\begin{eqnarray*}
W_{21}=W_{12}\frac{x_{t}(1-y_{t})}{y_{t}(1-x_{t})}=W_{12}r_{x_{t}y_{t}} \hspace{2.55mm} \mbox{and} \hspace{2.55mm} V_{12}=V_{21}\frac{y_{t}(1-x_{t})}{x_{t}(1-y_{t})}=V_{21}r_{y_{t}x_{t}}.
\end{eqnarray*}
 Therefore, the converted form of the system of equations is
\footnotesize
\begin{eqnarray}
x_{t+\Delta t}&=& \frac{W_{11}x_{t}y_{t}+2kW_{12}x_{t}(1-y_{t})}{W_{11}x_{t}y_{t}+W_{12}x_{t}(1-y_{t})+W_{21}y_{t}(1-x_{t})+W_{22}(1-x_{t})(1-y_{t})},   \nonumber \\
y_{t+\Delta t}&=&\frac{V_{11}x_{t}y_{t}+2\kappa V_{21}y_{t}(1-x_{t})}{V_{11}x_{t}y_{t}+V_{12}x_{t}(1-y_{t})+V_{21}y_{t}(1-x_{t})+V_{22}(1-x_{t})(1-y_{t})}. \label{eq:3}
\end{eqnarray} 
\normalsize
In order to transform equations (\ref{eq:3}) from discrete to continuous time, based on the concept of Mendelian segregation principle, we set $W_{ij}=1+\Delta t w_{ij}$, $V_{ij}=1+\Delta t v_{ij}$ $(i,j=1,2)$, $2kW_{12}=1+ 2\Delta t k w_{12}$, $2 \kappa V_{21}=1+2\Delta t \kappa v_{21}$ , and, assuming time between replication events much less than one, i.e., $\Delta t \ll 1$, the system of equations(\ref{eq:3}), can be approximated as \citep{Hofbauer1998,Traulsen2012}
\footnotesize
\begin{eqnarray}
x_{t+\Delta t}&=& x_{t} \frac{1+\Delta t(w_{11}y_{t}+2kw_{12}(1-y_{t})}{1+\Delta t (w_{11}x_{t}y_{t}+w_{12}x_{t}(1-y_{t})+w_{21}y_{t}(1-x_{t})+w_{22}(1-x_{t})(1-y_{t}))} \nonumber \\
 &\approx & x_{t}+\Delta t (w_{y_{t}}-<w>), \nonumber \\
y_{t+\Delta t}&= &y_{t} \frac{1+\Delta t(v_{11}x_{t}+2\kappa v_{21}(1-x_{t})}{1+\Delta t (v_{11}x_{t}y_{t}+v_{12}x_{t}(1-y_{t})+v_{21}y_{t}(1-x_{t})+v_{22}(1-x_{t})(1-y_{t}))} \nonumber \\
& \approx & y_{t}+\Delta t (v_{x_{t}}-<v>).
\end{eqnarray}
\normalsize 
Taking $\Delta t$ tending  to zero and dropping the index for discrete time, the time continuous evolutionary game dynamics (\ref{eq:1}) is recovered by
\begin{eqnarray}
\dot{x}&=&x (w_{y}-<w>),\nonumber \\
\dot{y}&=&y (v_{x}-<v> ) \label{eq:5}
\end{eqnarray}
where $w_{y}$ $= w_{11}y+2kw_{12}(1-y)$ and $v_{x}$ $= v_{11}x+2\kappa v_{21}(1-x)$ are the average fitnesses of the male and female $A_{1}$ alleles while $<w>$ $= w_{11}xy+2w_{12}x(1-y)+w_{22}(1-x)(1-y)$ and $<v>$ $ = v_{11}xy+2v_{21}y(1-x)+v_{22}(1-x)(1-y)$ are the average fitness in the male and female allele populations. And the payoff matrices for the interactions, describing the interactions of each member of the male allele population with every member of the female allele population and vice versa are 

\vspace{3mm}
\hspace{-6mm}
\scriptsize
\begin{tabular}{lcclllcc}
\multicolumn{3}{c}{Male population }& & &
\multicolumn{3}{c}{Female population}\\ \cline{1-3} \cline{6-8}
               & $A_{1}:y$ & $A_{2}:1-y$ & & & & $A_{1}:x$ & $A_{2}:1-x$ \\ \cline{1-3} \cline{6-8}
             $A_{1}:x$ & $w_{11}$ & $2kw_{12}$ & & & $A_{1}:y$ & $v_{11}$& $2\kappa v_{21}$  \\
             $A_{2}:1-x$ & $2(1-k)w_{12}r_{xy}$ & $w_{22}$& & & $A_{2}:1-y$& $2(1-\kappa)v_{21}r_{yx}$ & $v_{22}$  \\ \cline{1-3} \cline{6-8}                   
\end{tabular}  
\normalsize
\\  
\vspace{3mm}

In the overlapping generations, the non-constant viabilities of heterozygotes, $w_{21}$ and $v_{12}$, have no direct influence on the average fitnesses of male and female allele populations. Because of that, we shall call them pseudo fitnesses in the population fitness structure. Realistically, we can assume that the boundary values of the pseudo fitnesses on the boundary of the strategy region, $\{ 0 \leq x \leq 1, 0 \leq y \leq 1 \}$, are undefined or maximum or finite because $\frac{0}{0}$ is in-determinant: we can get anything at all from a $\frac{0}{0}$ situation. As the model is designed to think the interaction of two alleles at one locus of diploid genome and frequencies of alleles are optimizer of average fitness, the obtained selection dynamics builds the genotype-phenotype mapping structure through payoff matrices \citep{Traulsen2012,Haig2014}. However, we can not, here, say an evolutionary game i.e., frequency dependent selection, between alleles/haploids, is translated to the constant selection i.e., frequency-independent selection, in the diploids system because the pseudo fitnesses are non-constant.

\begin{figure}[t!]
\begin{center}
\includegraphics[width=1 \linewidth]{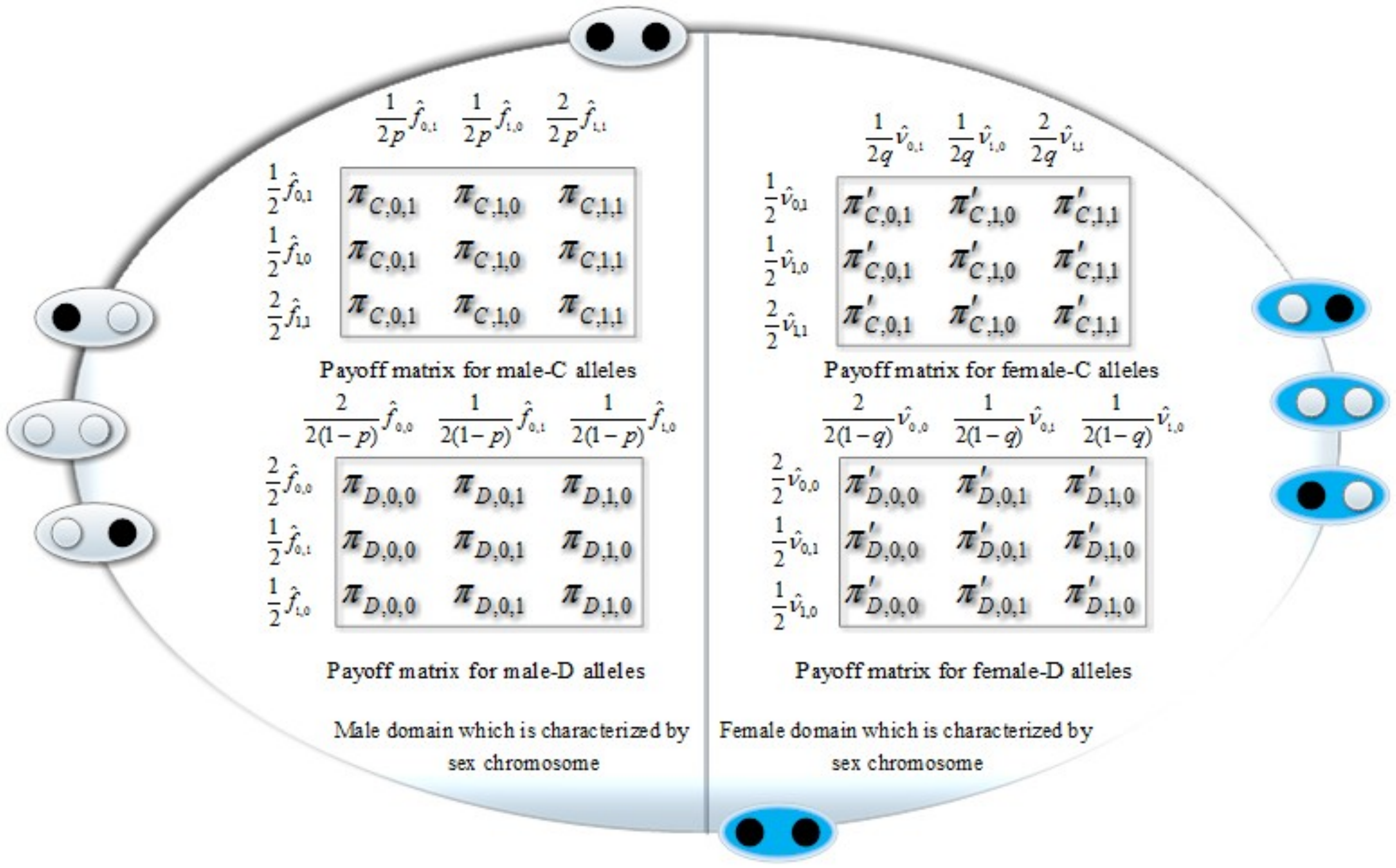}
\renewcommand{\figurename}{Fig.}
\caption[.]{\footnotesize {Evolutionary group selection between male allele population and female allele population separately. The four distinct groups of each sex  participate in the selection. Each group of each sex, separated by two different colors, is composed of two alleles whose phenotypic characters are characterized by  cooperating and defecting denoted by the light and dark circles. Proportions of $C$ alleles and $D$ alleles in the different configurations of the four groups in the male population and the female population have been shown along the left margins of the payoff matrices. And, along the top margins of matrices, frequencies of the impacts of different effective viabilities on the distinct groups are placed. In each group of males, the first entity (i.e., first allele) corresponds to the paternal origin while the second entity corresponds to the maternal origin whereas paternal and maternal entities appear in reverse order in the each group of females. It is noted that in male allelic pair $(A_{2},A_{1})$, effective viability of $A_{2}$ is to be  $2(1-k)w_{12}r_{pq}$ while effective viability of $A_{1}$ is $2kw_{12}r_{pq}$. And, in view of this observation, the payoff functions can be expressed in terms of effective viabilities (combined effect of viability and segregation ratio) in the following way: {\tiny $( \begin{array}{ccc} \pi_{C,0,1} &\pi_{C,1,0}&\pi_{C,1,1} \\ \pi_{D,0,0} & \pi_{D,0,1}& \pi_{D,1,0} \end{array})$ = $(\begin{array}{ccc} 2kw_{12}r_{pq} & 2kw_{12} & w_{11}\\ w_{22} & 2(1-k)w_{12}r_{pq} & 2(1-k)w_{12}\end{array})$} and {\tiny $(\begin{array}{ccc} \pi_{C,0,1}^{'} &\pi_{C,1,0}^{'}&\pi_{C,1,1}^{'} \\ \pi_{D,0,0}^{'} & \pi_{D,0,1}^{'}& \pi_{D,1,0}^{'} \end{array})$ = $(\begin{array}{ccc} 2\kappa v_{21}r_{qp} & 2\kappa v_{21} & v_{11}\\ v_{22} & 2(1-\kappa)v_{21}r_{qp} & 2(1-\kappa)v_{21}\end{array}) $}.}} \label{D Portrait}
\end{center}
\end{figure}

Next, let us turn to a group dynamics in perspective of group formation of male and female alleles. The concept introduces the non-random mating structure. We consider a group formation procedure through the games of Mendelian segregation and segregation distortion between any two allelic pairs of the four pairs (units), one of the alleles of each pair belonging to the male population and other allele to the female population, and allow any kind of group (pair) formation in respect of two strategies (those interpret phenotypic characters) --  $C$, defines the character of cooperating (attempting to cooperate) and $D$, links to the character of defecting (attempting to defect). As the genotype codes for the phenotype, to analyze the framework in genotypic flavor, here, we assume that the phenotypic character of cooperating is revealed by the allele $A_{1}$ while the allele $A_{2}$ causes the phenotypic character of defecting. The abundancy of $A_{1}$ shows the character of cooperating in the population. Our intention is to analyze how the frequency of $C$ allele is influenced in two cases of real interest where $k>\kappa$ and where $\kappa>k$. $f_{i,j}$ and $\nu_{j,i}$ are frequency of groups  in the male allele population and the female allele population respectively where $i=0,1$ and $j=0,1$, and each frequency term is defined as the frequency of the group with $(i+j)$ $C$ alleles and $2-(i+j)$ $D$ alleles in it. As the population states in the male allele population and the female allele population are to be characterized by $f=(f_{0,0},...,f_{1,1})$ and $\nu=(\nu_{0,0},...,\nu_{1,1})$ respectively, these frequencies have to satisfy the conditions \( \sum_{i=0}^{2} \sum_{j=0}^{2}f_{i,j}=1\) and \( \sum_{j=0}^{2} \sum_{i=0}^{2}\nu_{j,i}=1\) (compare with van Veelen \citep{Veelen2011}, Kerr and Godfrey-Smith \citep{Kerr2002}, Jensen and Rigos \citep{Jensen2012}); i.e., $f(t)$ belongs to 3-dimensional simplex $\Delta_{1}$ and it is defined on the trajectories of dynamical field as well as one to one corresponding with the population structure function $\hat{f}:[0,1]\times[0,1]\rightarrow \Delta_{1}$ while $\nu(t)$ is defined on the trajectories of dynamical field  on 3-dimensional simplex $\Delta_{2}$ through one to one corresponding with the population structure function $\hat{\nu}:[0,1]\times[0,1]\rightarrow \Delta_{2}$ (see \ref{A}). The associated viability selection structure of this non-random selection model has been explained and presented by the schematic diagram, Fig.\ref{D Portrait}. Hence, if the frequencies of male-$C$ alleles and female-$C$ alleles are denoted by $p$ and $q$ respectively, then the system of equations (\ref{eq:5}), becomes
\sloppy
\footnotesize
\begin{eqnarray}
 \dot{p}\hspace{-1mm}&=&\hspace{-1mm}p\Bigl( \frac{1}{2p}\sum_{i=0}^{1}\sum_{j=0}^{1}(i+j)\hat{f}_{i,j}(p,q)\pi_{C,i,j}-\Bigl[ \frac{p}{2p}\sum_{i=0}^{1}\sum_{j=0}^{1}(i+j)\hat{f}_{i,j}(p,q)\pi_{C,i,j} \nonumber \\ & &+ \frac{1-p}{2(1-p)}\sum_{i=0}^{1}\sum_{j=0}^{1}(2-(i+j))\hat{f}_{i,j}(p,q)\pi_{D,i,j}\Bigr]  \Bigr), \nonumber \\
\dot{q}\hspace{-1mm}&=&\hspace{-1mm}q\Bigl(\frac{1}{2q}\sum_{j=0}^{1}\sum_{i=0}^{1}(i+j)\hat{\nu}_{j,i}(q,p)\pi_{C,j,i}^{'}-\Bigl[ \frac{q}{2q}\sum_{j=0}^{1}\sum_{i=0}^{1}(i+j)\hat{\nu}_{j,i}(q,p)\pi_{C,j,i}^{'} \nonumber \\ & &+\frac{1-q}{2(1-q)}\sum_{j=0}^{1}\sum_{i=0}^{1}(2-(i+j))\hat{\nu}_{j,i}(q,p)\pi_{D,j,i}^{'}\Bigr] \Bigr). \label{eq:6}
\end{eqnarray}
\normalsize
To ensure the existence of solutions and uniqueness of solutions, we will assume that $\hat{f}$ and $\hat{\nu}$ are Lipschitz continuous such that $f(t)=\hat{f}(p(t),q(t))$ and $\nu(t)=\hat{\nu}(q(t),p(t))$.

It is important to note that in any selection dynamics, genic and genotypic analyses are both valid but processes of the analyses follow up in the different ways. In random mating population, the genic fitness of male allele $A_{1}$ relative to male allele $A_{2}$ is $\frac{w_{1}}{w_{2}}=\frac{xw_{11}+(1-x)w_{12}}{xw_{12}+(1-x)w_{22}}$. The equilibrium frequency $x^{\ast}$ can be found by setting $w_{1}=w_{2}$ (see Haig \citep{Haig2014} and reference therein). In the following sections, the evolution frameworks are analyzed in perspectives of evolutionary game theory as well as group selection under the common configuration arrangement of associated biological parameters.

\section{Existence and stability of polymorphic equilibrium}

\subsection{Population genetics in evolutionary game theory framework}
Mathematically, an equilibrium point $(x^{\ast},y^{\ast})$, is a point that satisfies the system of equations $\{\dot{x}=0, \dot{y}=0\}$ and if a solution starts at this point, it remains there forever; i.e., at equilibrium, allele frequency remains unaltered overtime. According to the structural framework, here, natures of equilibrium points are classified into the two categories -- trivial and nontrivial where in the trivial equilibrium point either $A_1$ or $A_2$ is absent.

At the equilibrium point $( x^{\ast}, y^{\ast})$, the system of equations (\ref{eq:5}), can be written in the following way,
{\footnotesize
\begin{eqnarray}
\hspace{-2.5mm} x^{\ast}(1-x^{\ast})\Bigl[\Bigl( w_{11}-2(1-k)w_{12}\frac{x^{\ast}(1-y^{\ast})}{y^{\ast}(1-x^{\ast})}\Bigr)y^{\ast}+\Bigl(2 k w_{12}-w_{22} \Bigr)(1-y^{\ast}) \Bigr]&=&0, \nonumber \\
y^{\ast}(1-y^{\ast})\Bigl[\Bigl( v_{11}-2(1-\kappa)v_{21}\frac{y^{\ast}(1-x^{\ast})}{x^{\ast}(1-y^{\ast})}\Bigr)x^{\ast}+\Bigl(2 \kappa v_{21}-v_{22} \Bigr)(1-x^{\ast}) \Bigr]&=&0 \label{eq:7}.
\end{eqnarray}}
\normalsize
\noindent Therefore, the trivial equilibrium can only be calculated at $w_{ij}=v_{ji}=0 \hspace{0.6mm} (i \neq~ j)$ and the equilibrium points are 
$E_{0}=(0,0)$, $E_{1}=(0,1)$, $E_{2}=(1,0)$, $E_{3}=(1,1)$, and $E_{4}=(\frac{v_{22}}{v_{11}+v_{22}},\frac{w_{22}}{w_{11}+w_{22}} )$ where the last one is nontrivial equilibrium. Exclude the constraint $w_{ij}=v_{ji}=0 \hspace{.4mm} (i \neq~ j)$, and assuming there are no trivial equilibria, the system of equations (\ref{eq:7}), is expressed as,
\begin{eqnarray}
a_{1}x^{\ast}+a_{2}y^{\ast}-a_{3}x^{\ast}y^{\ast}&=&a_{4}, \nonumber \\
a_{5}x^{\ast}+a_{6}y^{\ast}-a_{7}x^{\ast}y^{\ast}&=&a_{8} \label{eq:8}
\end{eqnarray} 
where $a_{1}=w_{22}-2w_{12}$, $a_{2}=w_{11}-2kw_{12}+w_{22}$, $a_{3}=w_{11}-2w_{12}+w_{22}$, $a_{4}=w_{22}-2 k w_{12}$, $a_{5}=v_{11}-2\kappa v_{21}+v_{22}$, $a_{6}=v_{22}-2v_{21}$, $a_{7}=v_{11}-2v_{21}+v_{22}$ and $a_{8}=v_{22}-2 \kappa v_{21}$. Hence, as in the system of equations (\ref{eq:8}), $( x^{\ast}, y^{\ast}) \neq (1,1)$, there exists only one nontrivial equilibrium, which denotes by $E_{5}$. Presently, for the sake of analysis, we introduce some special terminologies  associated with  heterozygote effect, e.g., the additive selections refer to the fitness effects which are additive, the recessive selections refer to the fitness effects which are recessive ; and restrict ourself around the polymorphic equilibrium $E_{5}$.
\hbadness=10000 
\begin{figure}[t!]
\makebox[\textwidth][c]{
\begin{minipage}{0.4\textwidth}
\centering
First Column \\[10pt]
\includegraphics[width=1.2\textwidth]{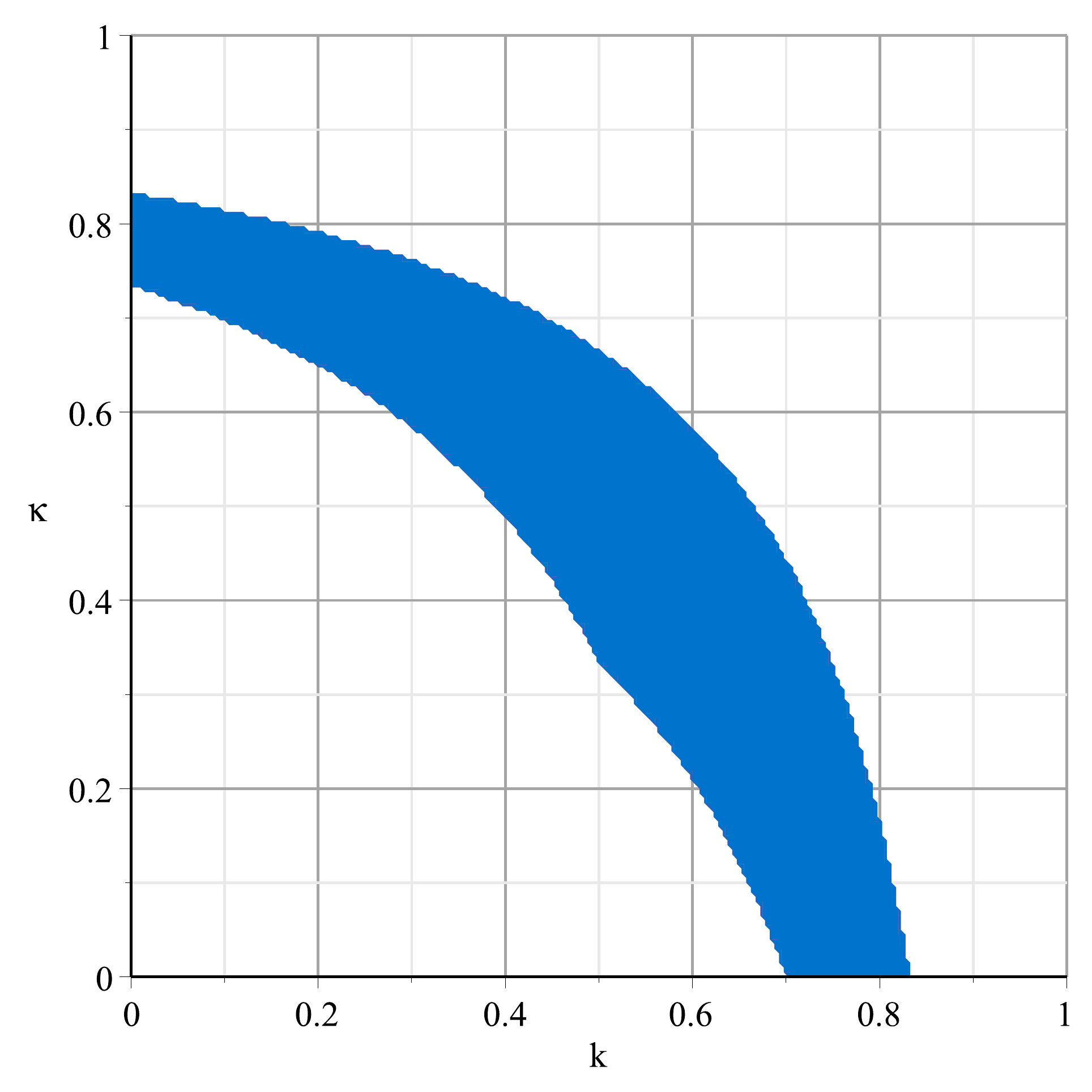}
\label{fig:test1}
\end{minipage}%
\hspace{1cm}
\begin{minipage}{.4\textwidth}
\centering
Second Column \\[10pt]
\includegraphics[width=1.2\textwidth]{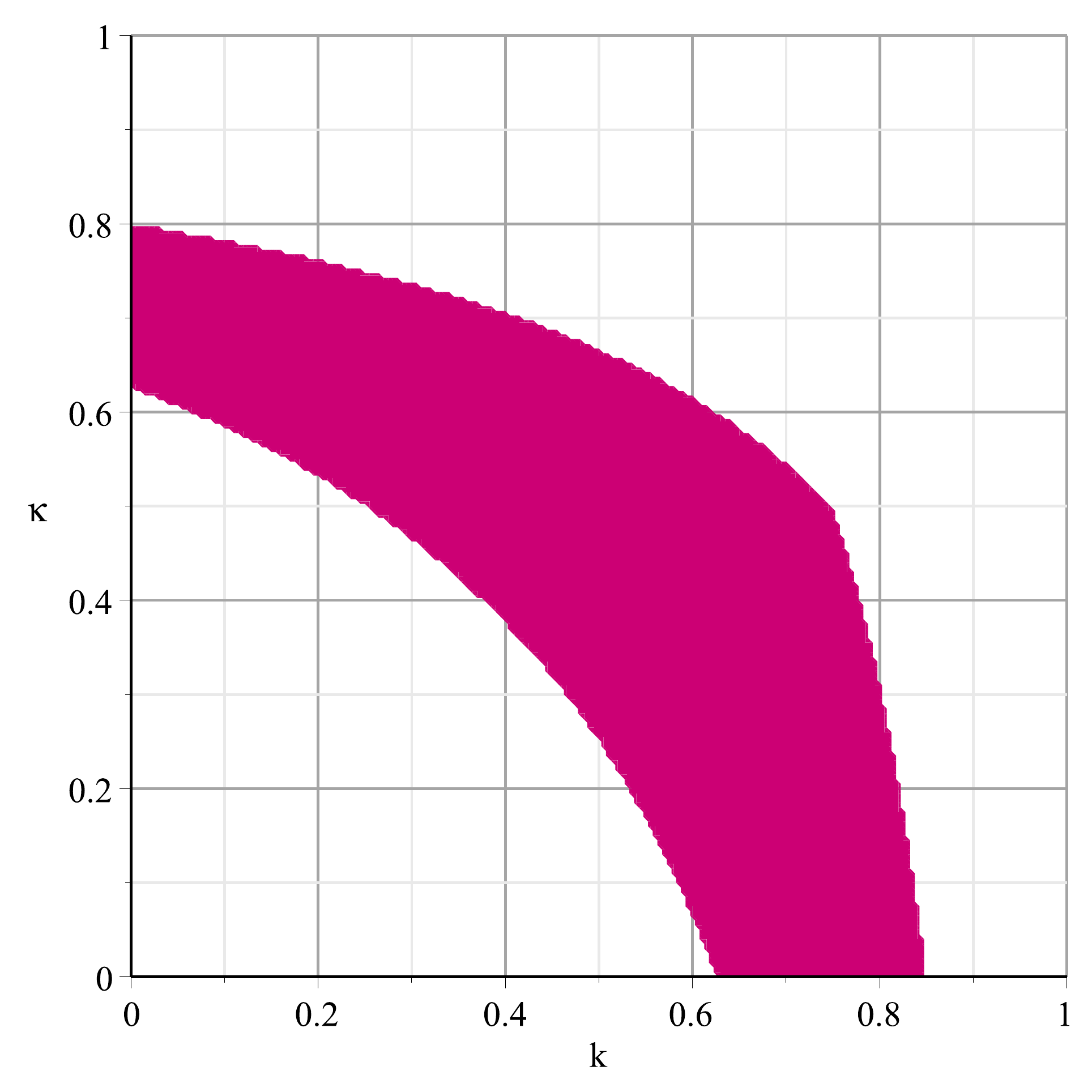}
\label{fig:test2}
\end{minipage}
\hspace{1cm}
\begin{minipage}{.4\linewidth}
\centering
Third Column \\[10pt]
\includegraphics[width=1.2\linewidth]{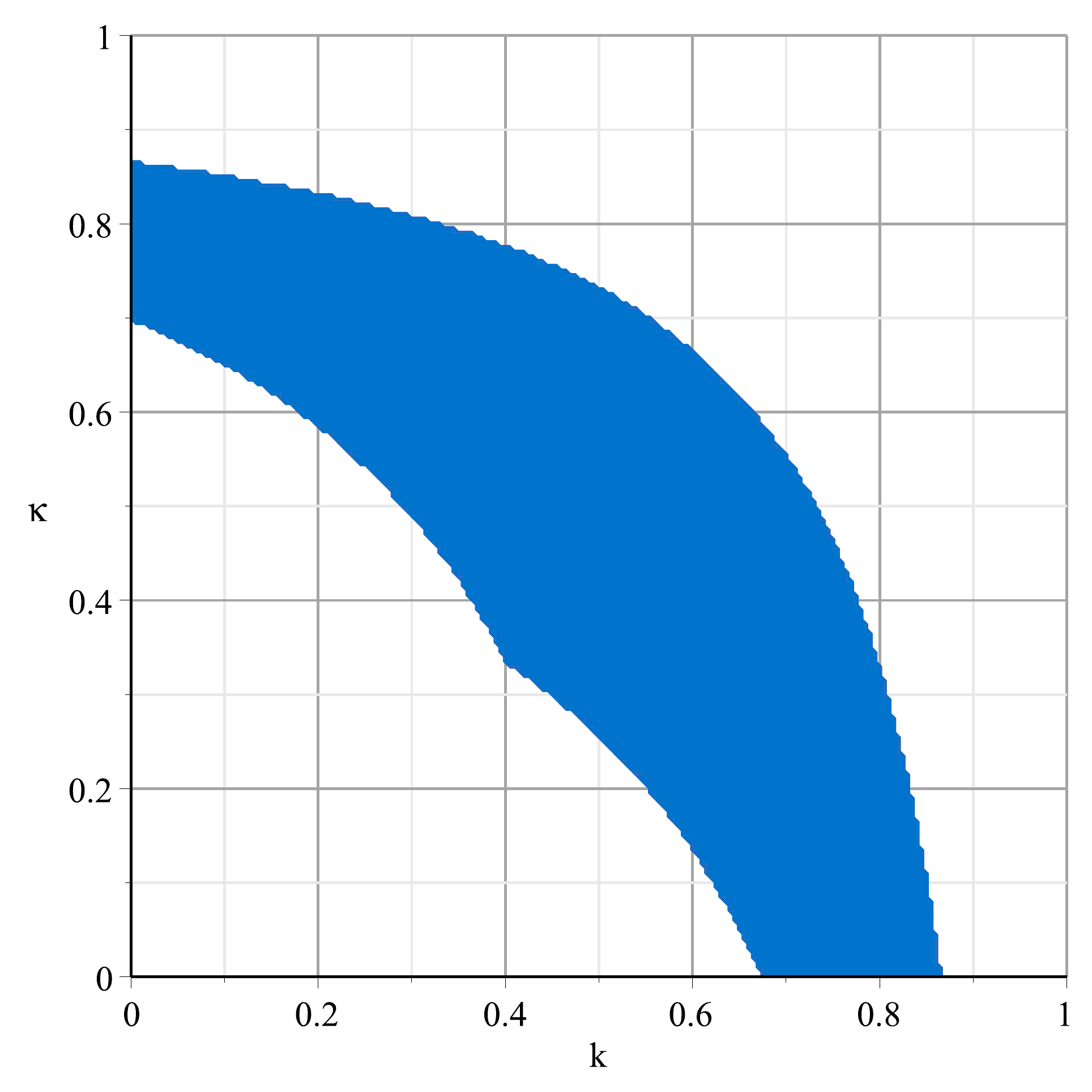}
\label{fig:test3}
\end{minipage}}\\
\makebox[\textwidth][c]{
\begin{minipage}{.4\linewidth}
\centering
\includegraphics[width=1.2\linewidth]{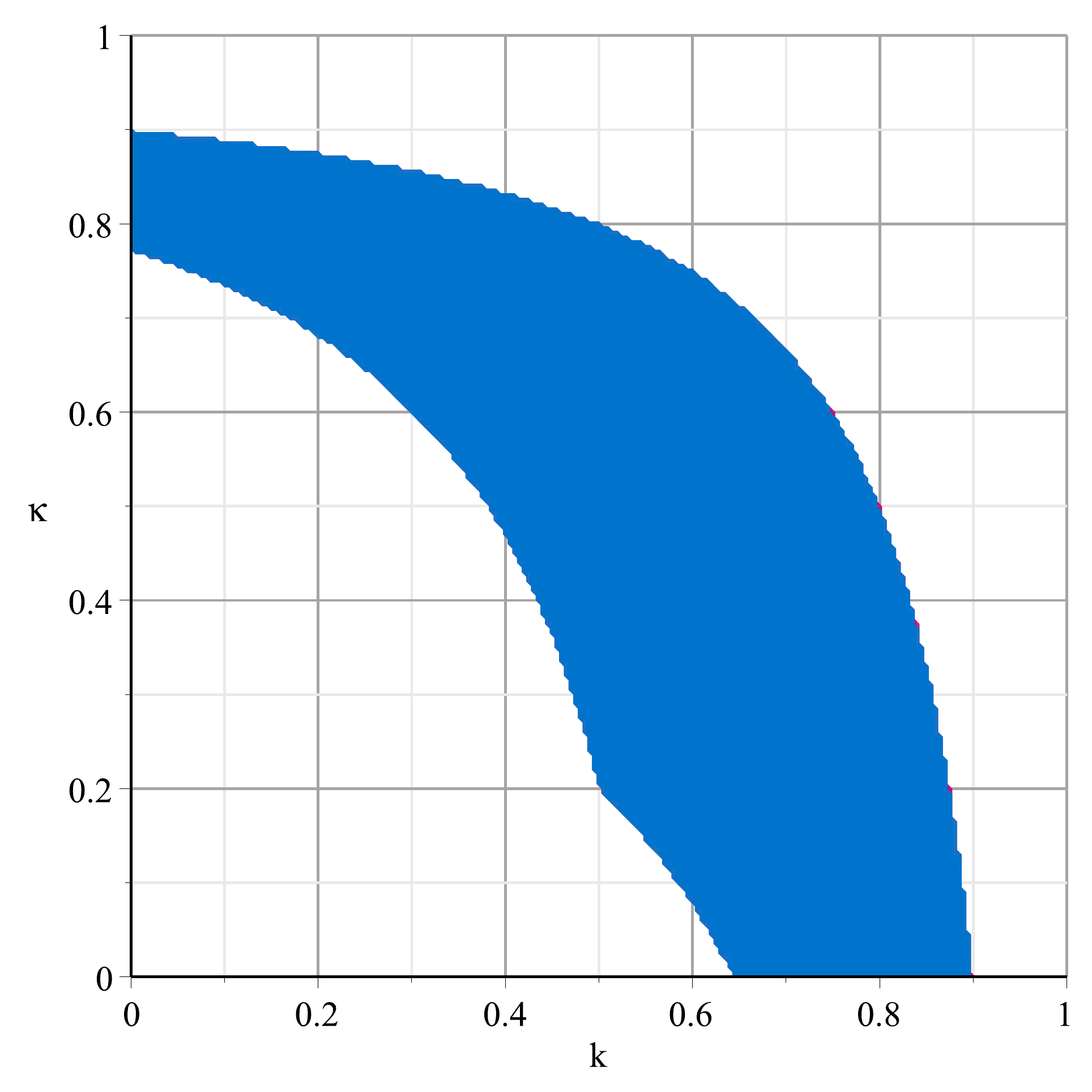}
\label{fig:test4}
\end{minipage}%
\hspace{1cm}
\begin{minipage}{.4\linewidth}
\centering
\includegraphics[width=1.2\linewidth]{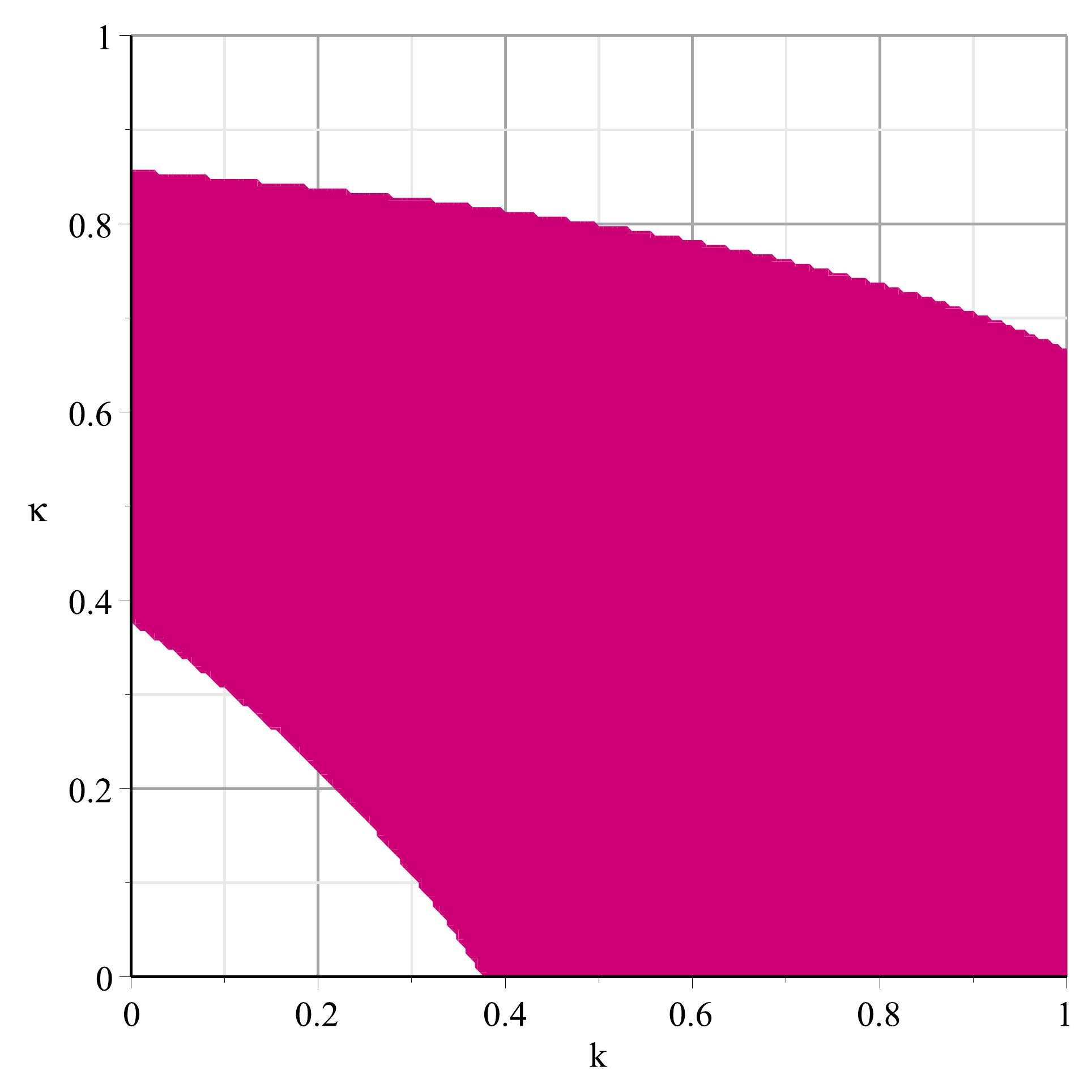}
\label{fig:test5}
\end{minipage}
\hspace{1cm}
\begin{minipage}{.4\linewidth}
\centering
\includegraphics[width=1.2\linewidth]{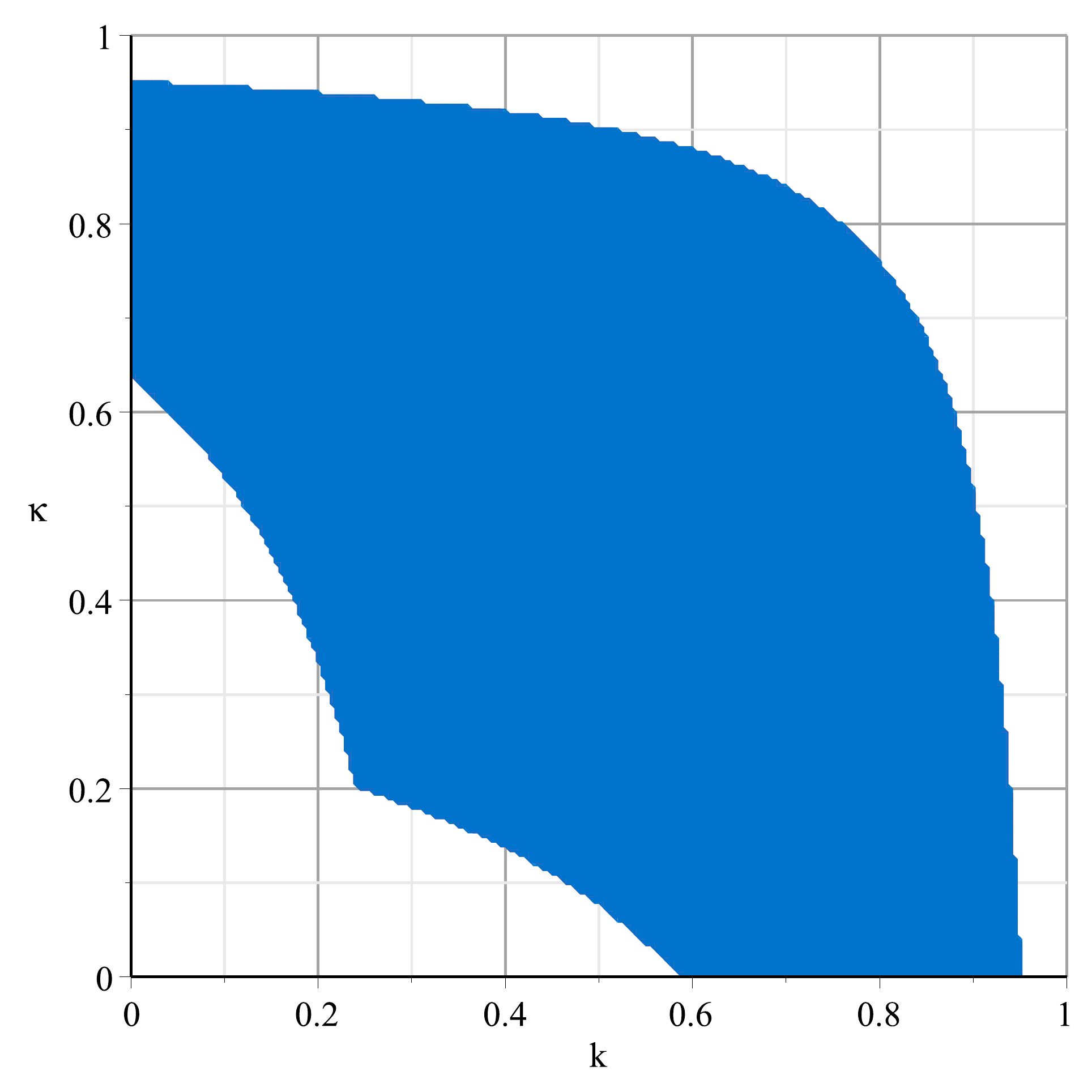}
\label{fig:test6}
\end{minipage}} 
\renewcommand{\figurename}{Fig.}
\caption[.]{\footnotesize {Array representation of viability effect on the existence and stability conditions of the internal equilibrium. We consider different alleles are favored in the two sexes where $w_{21}$ and $v_{12}$ are pseudo fitnesses (two variable quantities) of the population fitness structure. The effects of the dominance fitnesses are presented along the first column in which the fitness ranges being {\tiny $( \begin{array}{rl} 1 & 1.5:1:2.5 \\ w_{21} & 1.5:1:2.5 \end{array}) $} and {\tiny $( \begin{array}{rl} 1.5:1:2.5 & v_{12} \\ 1.5:1:2.5 & 1 \end{array}) $}. In the second and third columns the outcomes of the recessive and overdominant selection effects have been plotted respectively, where the fitness ranges are {\tiny $( \begin{array}{rl} 1 & 1 \\ w_{21} & 1.5:1:2.5 \end{array}) $} and {\tiny $( \begin{array}{rl} 1.5:1:2.5 & v_{12} \\ 1 & 1 \end{array}) $} in the second column, and in the third column the fitness ranges are {\tiny $( \begin{array}{rl} 1 & 1.5:1:2.5 \\ w_{21} & 1.2 \end{array}) $} and {\tiny $( \begin{array}{rl} 1.2 & v_{12} \\ 1.5:1:2.5 & 1 \end{array}) $}. Sex-specific fitness arrays having linear increment of components are one to one correspondence with each other. Spring 1 colored shaded region in $(k,\kappa)$  space yields one stable equilibrium while one unstable equilibrium is to be found in Spring 2 colored shaded region.}}\label{fig:stability}
\end{figure}

Comparative study on the numerical data structure reveals that, except the additive selection, in the cases of dominant, recessive and overdominant selections, non trivial equilibrium $E_{5}$ exists where only in the recessive selection nature of the equilibrium point is unstable (saddle point). As at the trivial equilibrium points, pseudo fitnesses $w_{21}$ and $v_{12}$ are undefined, it is beyond the scope of this article to verify whether Prout Criterion \citep{Prout1968} is applicable to find out the nature of polymorphic equilibrium, $E_{5}$. The possibility of arising the stable polymorphic equilibrium is high in the overdominance structure which is a well-known result having been shown in Fig.\ref{fig:stability} where either $\kappa+k=1$ or $\kappa=k$ or both relations are held in a wide range in comparison to the dominant viability effect. To plot the figures we have considered that different alleles are favored in the two sexes. If, in spite of different alleles, same alleles are favored in the two sexes, then it can be shown that the possibility of existence of equilibrium at the Mendelian segregation is low where in most of the cases the Mendelian segregation point becomes a boundary point of the stable or unstable regions \citep{Kidwell1977,Ewens2004}. Here, it is noticeable observation that when alleles $A_{1}$ are favored in the both sexes, the shaded region relating to feasible solution in $(k,\kappa)$ space tries to spread toward the $(0,0)$ corner and on the contrary when the both sexes favor alleles $A_{2}$, the shaded region tries to spread toward the opposite corner, $(1,1)$. This observation has a simple biological interpretation. At the equilibrium circumstance, the abundant presence of allele $A_{1}$ means that allele $A_{2}$ will be transmitted to a fraction $(1-k)$ of its bearers' sperm and a fraction $(1-\kappa)$ of its bearers' eggs where the transmitted rate of $A_{2}$ is higher than that of $A_{1}$ and sperm will fertilize eggs in such a way that $A_{2}$ can invade a gene pool near fixation for $A_{1}$. On the other hand, at the equilibrium circumstance, the abundant presence of allele $A_{2}$ means that allele $A_{1}$ will be transmitted to a fraction $k$ of its bearers' sperm and a fraction $\kappa$ of its bearers' eggs where the transmitted rate of $A_{1}$ is higher than that of $A_{2}$ and sperm will fertilize eggs in such a way that $A_{1}$ can invade a gene pool near fixation for $A_{2}$ \citep{Ubeda2004}.

\subsection{ Population genetics in group selection framework}
Without loss of generality, to find out the effects of segregation ratios $(k,\kappa)$ and mating parameter, $\alpha$, on the frequencies of male-C alleles and female-C alleles, we fix the following set of viabilities in both sexes: $(1, s, 1.5)$ where $s$ is the fitness of heterozygote, and mating parameter, $\alpha$ has been defined in \ref{B}. The mating parameter is a scale parameter of the population structure of mixture of random grouping and clonal interaction where a population state is characterized by the frequencies of four allelic pairs (a group state). The scale parameter calibrates the non-randomness, ranging from $0$ to $1$. Value $\alpha=0$ marks  the random mating while the clonal interaction character is indicated by the value of $\alpha=1$. 

Considering equations (\ref{eq:6}), it is easy to show that, under any population structure, an interior state $(p^{\ast}, q^{\ast})$ is an equilibrium if and only if: $\pi_{C}(p^{\ast}, q^{\ast})=\pi_{D}(p^{\ast}, q^{\ast})$ and $\pi_{C}^{'}(q^{\ast}, p^{\ast})=\pi_{D}^{'}(q^{\ast}, p^{\ast})$ simultaneously hold, i.e.,
\begin{eqnarray}
b_{1}p^{\ast}+b_{2}q^{\ast}+b_{3}p^{\ast}q^{\ast}&=&b_{4}, \nonumber \\
b_{5}p^{\ast}+b_{6}q^{\ast}+b_{7}p^{\ast}q^{\ast}&=&b_{8} \label{3.3}
\end{eqnarray} 
where $b_{1}=\alpha+2(1-\alpha)s-1.5$, $b_{2}=(1-\alpha)(2ks-2.5)$, $b_{3}=(1-\alpha)(2.5-2s)$, $b_{4}= \alpha+2(1-\alpha)ks-1.5$, $b_{5}=(1-\alpha)(2\kappa s-2.5)$, $b_{6}=\alpha+2(1-\alpha)s-1.5$, $b_{7}=(1-\alpha)(2.5-2s)$ and $b_{8}=\alpha+2(1-\alpha)\kappa s-1.5$. Similar to the previous subsection, we assume that there are no trivial equilibria, and as, consequence of that assumption, in equations \ref{3.3} $( p^{\ast}, q^{\ast}) \neq (1,1)$, there exists only nontrivial equilibrium $E_{6}$.

\begin{figure}[b!]
\makebox[\textwidth][c]{
\begin{minipage}{.4\linewidth}
\centering
\includegraphics[width=1.2\linewidth]{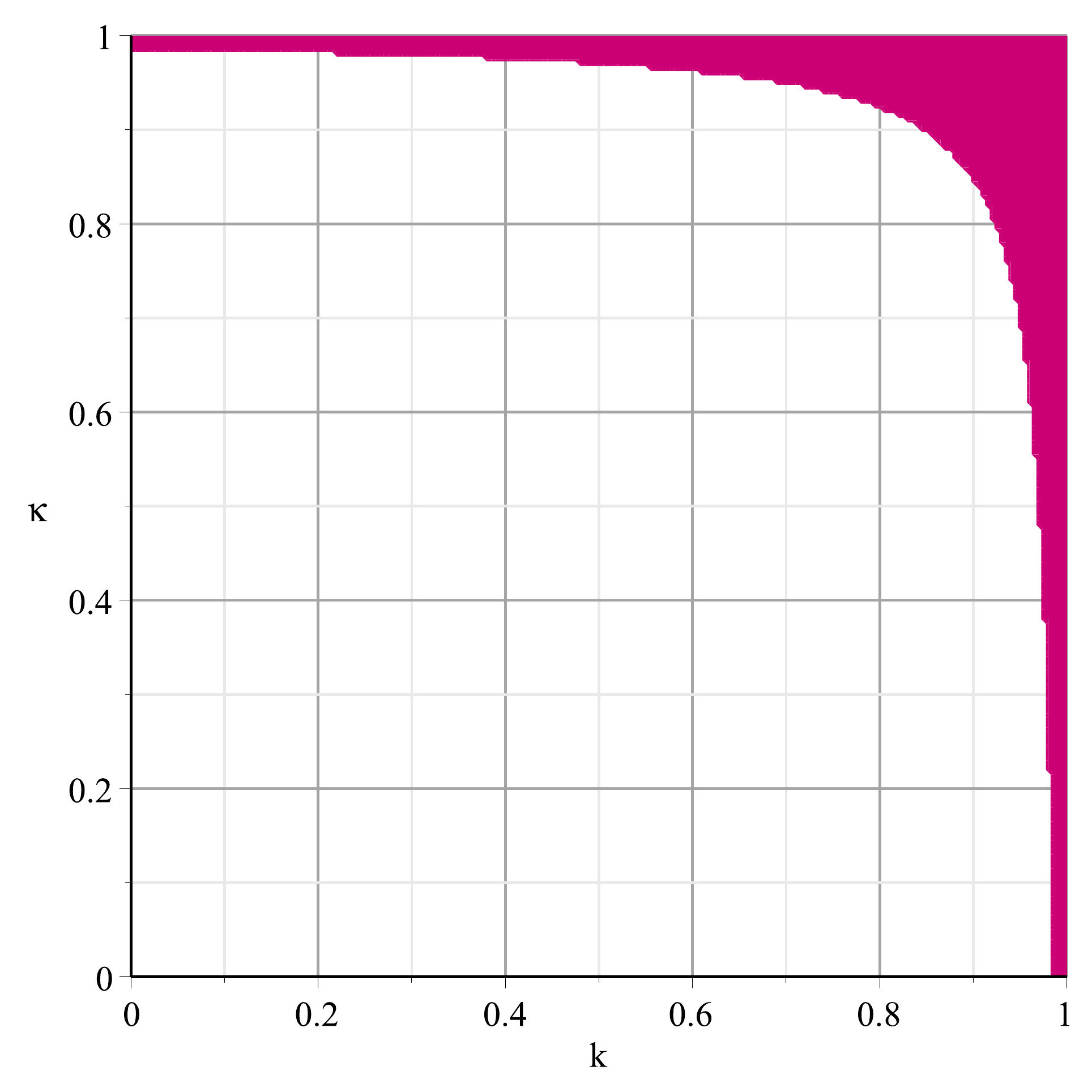}
\label{fig:test101}
\end{minipage}%
\hspace{1cm}
\begin{minipage}{.4\linewidth}
\centering
\includegraphics[width=1.2\linewidth]{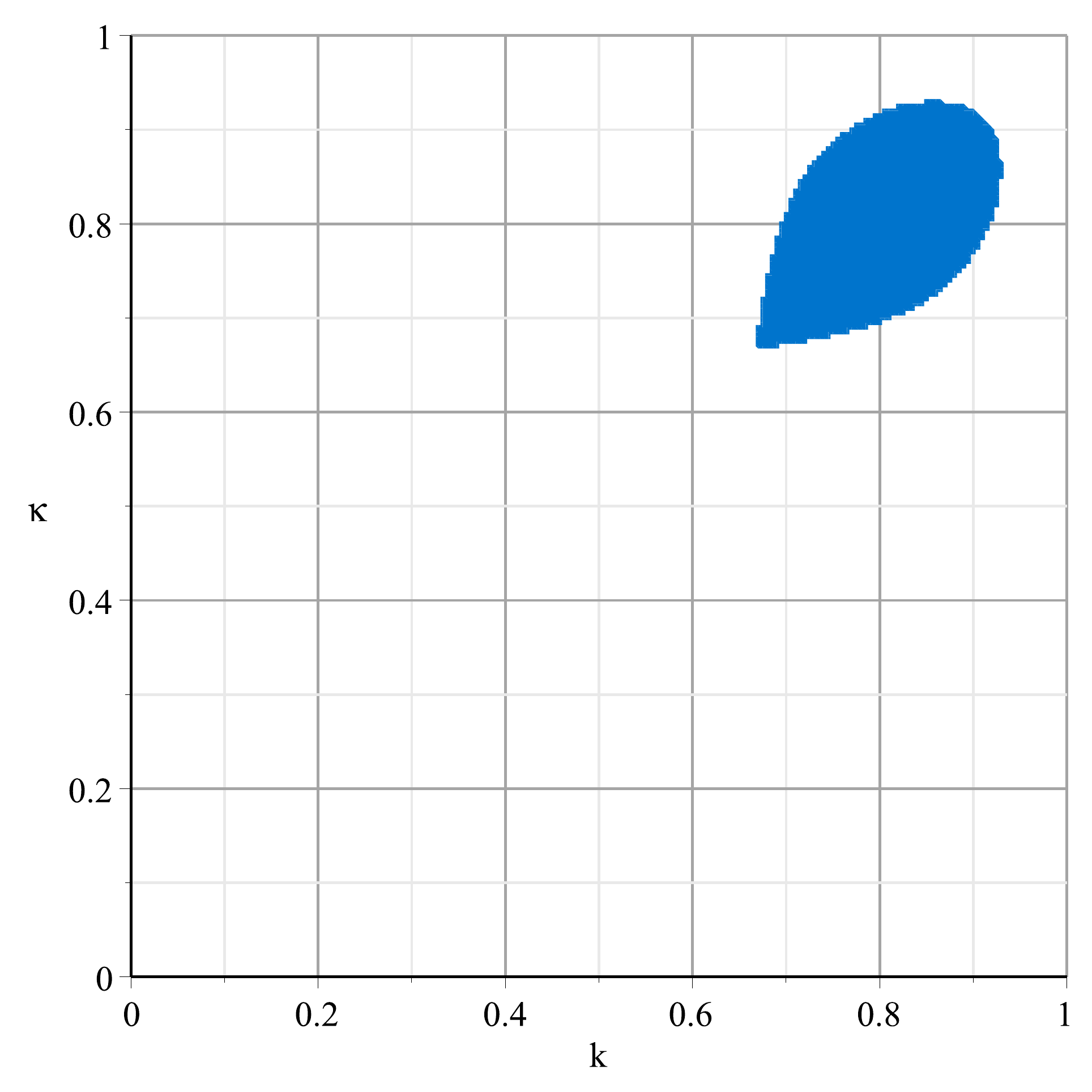}
\label{fig:test102}
\end{minipage}
\hspace{1cm}
\begin{minipage}{.4\linewidth}
\centering
\includegraphics[width=1.2\linewidth]{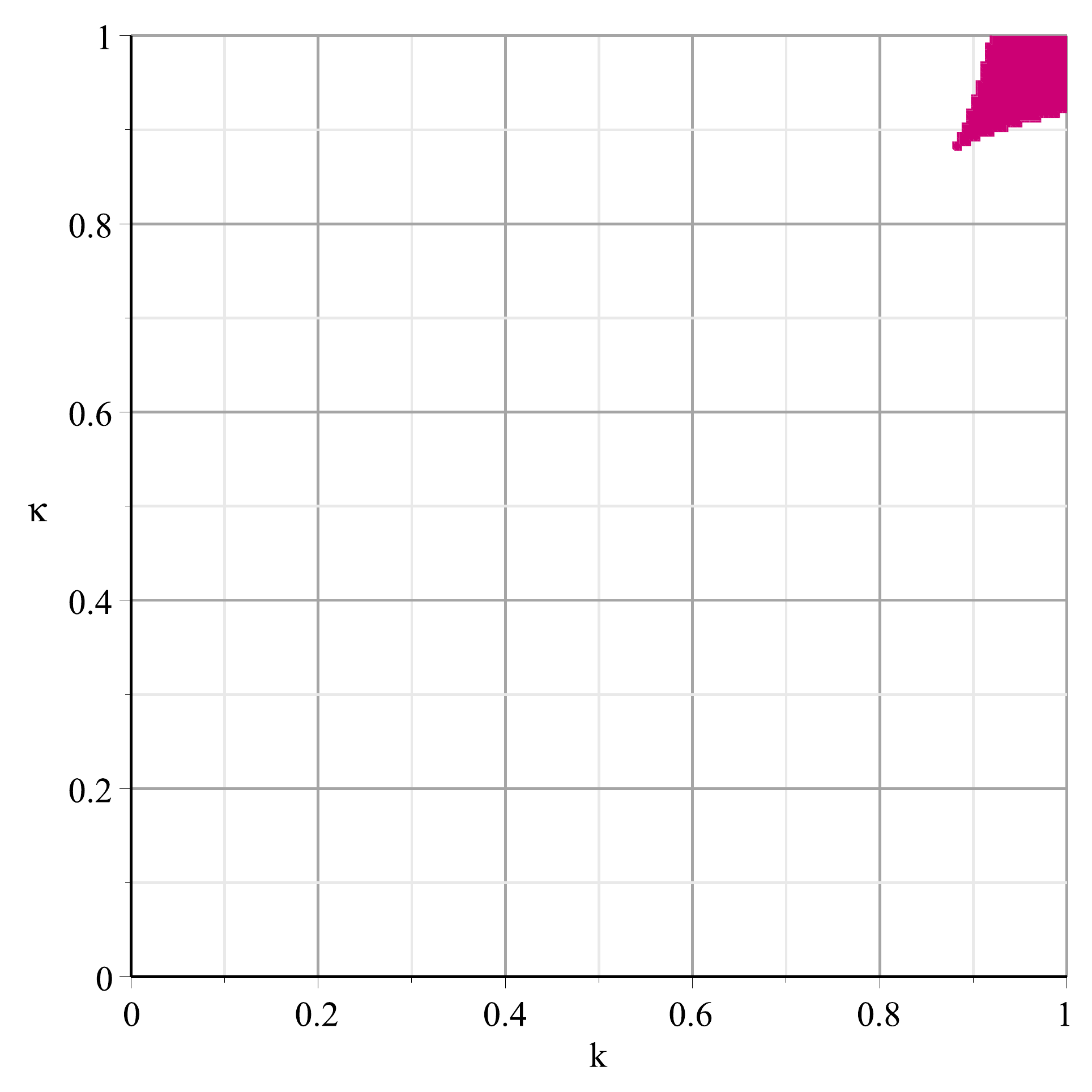}
\label{fig:test103}
\end{minipage}} 
\renewcommand{\figurename}{Fig.}
\caption[.]{\footnotesize {Row representation of viability effect on the existence and stability conditions of the internal equilibrium. The first figure corresponds to the recessive selection of $\alpha=0.6$ whereas the second and third figures, in the common value of $s=2$ of overdominant selection, are plotted  with values of $\alpha=0.7$ and $\alpha=0.8$ respectively. Color interpretations are same as in Fig.\ref{fig:stability}}.}\label{fig:GSstability}
\end{figure}

Here in this subsection too, as the close form of the root $E_{6}$ is sufficiently complicated as well as long and as it has no insight other than the algebraic logic, in order to analyze the group selection framework we give emphasis on the data analysis. And the most significant observation is that the possibility of existence of internal equilibrium, $E_{6}$ is low in $(k, \kappa)$ space when mating nature is becoming non-random. Moreover, at the point of complete segregation (clonal interaction) the internal equilibrium is disappeared. However, we can not expect that the clonal interaction and underdominant selection with $w_{i,j}=v_{j,i}=0 \hspace{.4mm} (i \neq~ j) $ would give rise to analogous dynamics  because we have seen in the previous subsection  that the underdominant selection presents a dynamics with internal equilibrium. Later, the nature of internal  equilibrium in the underdominant selection will be examined. Similar to the random mating, here internal equilibrium  is also absent in the additive selection.

Stable nature is bifurcated at a fixed value of the mating parameter $\alpha$ of the overdominant selection that being shown in Fig.\ref{fig:GSstability}. Exploring the sequence of figures, with $s=2$, in the range $0\leq \alpha \leq 0.7$, a stable nature of decreasing region in $(k, \kappa)$ space, which tends  towards $(1,1)$ corner is noticed and this stable region becomes unstable for further increasing in $\alpha$ past $0.7$. Manually, one can find out the  bifurcation value of $\alpha$.  However, the plot configuration strongly suggests that the bifurcation value of $\alpha$ is directly influenced by heterozygote fitness value $s$ and this is why, with higher value of $s$ the stable nature of decreasing region in $(k, \kappa)$ space remains unaltered throughout the range $0\leq \alpha < 1$. On the other hand, in the range $0\leq \alpha \leq 0.7$ of the dominant selection and in the range $0\leq \alpha \leq 0.6$ of the recessive selection the obtained feasible regions in $(k, \kappa)$ space are stable and unstable respectively. Out side of these ranges, the regions in $(k, \kappa)$ space become the regions connected to the non feasible solutions.

\section{The model analysis}
\subsection[Population genetics in evolutionary game theory framework]{\texorpdfstring{Population genetics in evolutionary \\ game theory framework}{Population genetics in evolutionary game theory framework}}

\begin{wrapfigure}{r}{0.5\textwidth}
\vspace{-99pt}
\begin{center}
\includegraphics[width=.47\textwidth]{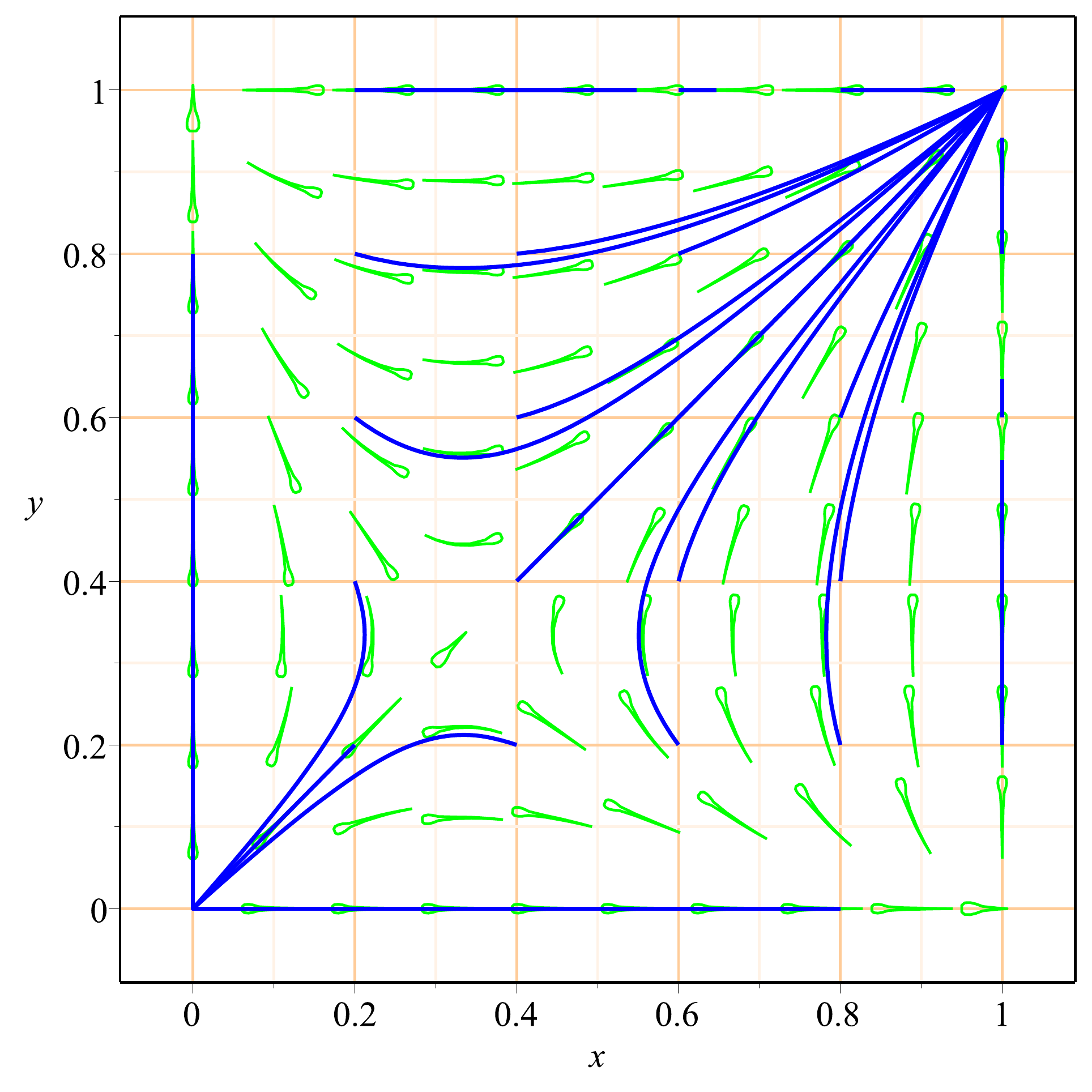}
\end{center}
\vspace{-20pt}
\renewcommand{\figurename}{Fig.}
\caption{\footnotesize {Underdominant selection: favoring the same alleles in the two sexes}}
\vspace{-18pt}
\label{Phase Portrait}
\end{wrapfigure}

Examples of heterozygote inferiority (underdominance) in nature are relatively few, unless a species lives in a coarse-grained environment where there is strong selection for alternative homozygous genotypes. However, to assess the strategies involving chromosome translocations in genetic control, the models of underdominance are useful \citep{Hedrick2005}. 

In the present model, heterozygote inferiority yields five equilibria. Two of them, the equilibrium points $E_{0}$ and $E_{3}$, are stable and size of a basin of attraction is dependent on the viability strength, which is revealed in Fig.\ref{Phase Portrait}, plotted with the set value $((w_{11}=2,w_{22}=1),(v_{11}=2,v_{22}=1))$, although if different alleles are favored in the two sexes under the condition, $w_{22}=v_{11}$ or $w_{11}=v_{22}$, then the obtained basins of attraction are of equal size, and in the sense, $x^{\ast}+y^{\ast}=1$, the polymorphic equilibrium $E_{4}$  changes the position along the principal diagonal with the variation of viability parameters. 

In conventional procedure of evolutionary game theory, the underdominance viabilities (payoffs) are presented as the outcomes of the bimatrix game:

\vspace{3mm}
\hspace{-6mm}
\scriptsize
\begin{tabular}{lcclllcc}
\multicolumn{3}{c}{Male population }& & &
\multicolumn{3}{c}{Female population}\\ \cline{1-3} \cline{6-8}
               & $A_{1}:y$ & $A_{2}:1-y$ & & & & $A_{1}:x$ & $A_{2}:1-x$ \\ \cline{1-3} \cline{6-8}
             $A_{1}:x$ & $w_{11}$ & $0$ & & & $A_{1}:y$ & $v_{11}$& $0$  \\
             $A_{2}:1-x$ & $0$ & $w_{22}$& & & $A_{2}:1-y$& $0$ & $v_{22}$  \\ \cline{1-3} \cline{6-8}                   
\end{tabular}  
\normalsize
 
\vspace{4mm}

\noindent The game has two Nash equilibrium points in pure strategies, namely $(A_{1},A_{1})$, $(A_{2},A_{2})$, and  another Nash equilibrium point in mixed strategies $((\frac{v_{22}}{v_{11}+v_{22}},\frac{v_{11}}{v_{11}+v_{22}}),(\frac{w_{22}}{w_{11}+w_{22}},\frac{w_{11}}{w_{11}+w_{22}}))$, with the corresponding values of expected payoffs (average fitnesses),
\[(<w>,<v>)=(\pi_{1}(X^{\ast},Y^{\ast})=\frac{w_{11}w_{22}}{w_{11}+w_{22}},\pi_{2}(Y^{\ast},X^{\ast})=\frac{v_{11}v_{22}}{v_{11}+v_{22}}).\]
From this standpoint, we can conclude that the population mean fitness of both sexes is maximized at the dynamical equilibrium point over those variables that the allele controls, under the assumption that the other allele's variables of the opposite sex are held fixed. However, as $\pi_{1}(e^{i},e^{i})>\pi_{1}(X^{\ast},Y^{\ast})$, $\pi_{2}(e^{i},e^{i})>\pi_{2}(Y^{\ast},X^{\ast})$ $(i=1,2)$ and as, according to the Fundamental Theorem of Natural Selection, fitness is increased or remains the same over time (in other words, the natural selection does not allow a population to decrease in average fitness), thus it is clearly expected in the evolutionary game  point of view that the Nash equilibrium point in mixed strategies, $E_{4}$, is unstable (see Fig.\ref{Phase Portrait}). The population mean fitness in both sexes is maximized at either $E_{0} \equiv (e^{2},e^{2})$ or $E_{3} \equiv (e^{1},e^{1})$ and therefore, each of the two Nash equilibrium points in pure strategies corresponds to $ESS$. It is to be noted that in the present subsection and henceforth the analysis structure will mainly be based on the combining result of the three propositions:


\newdefinition{pro}{Proposition}[section]
\begin{pro}
 The system of equations(\ref{eq:5}), is a regular, aggregate monotonic selection dynamics (referring to Samuelson and Zhang \cite{Samuelson1992}).
\end{pro}
\begin{pro}
The bimatrix game associated with the system of equations(\ref{eq:5}), follows the Fundamental Theorem of Natural Selection (referring to Weibull \cite{Weibull1997}). \label{p4.2}
\end{pro}
\begin{pro}
The asymmetric two-population game, adopts the follwing inclusion relation (referring to Accinelli and Sanchez Carrera \cite{Accinelli2011}):
\[ ESSset \subseteq \mbox{asymtotically stable set} \subseteq NEset \subseteq \mbox{equilibrium solution set} \]
\end{pro}

We know that the selection dynamics (\ref{eq:5}) is aggregate monotonic if for all $X^{\prime},X^{\prime \prime} \in \Delta_{1}$, $\pi_{1}(X^{\prime},Y)>\pi_{1}(X^{\prime \prime},Y)$ $\Rightarrow$ $\sum_{i=1\mapsto y}^{i=2\mapsto 1-y}(x_{i}^{\prime}-x_{i}^{\prime \prime})(w_{y}-<w>)>0$ and for all $Y^{\prime},Y^{\prime \prime} \in \Delta_{2}$, $\pi_{2}(Y^{\prime},X)>\pi_{2}(Y^{\prime \prime},X)$ $\Rightarrow$ $\sum_{i=1\mapsto x}^{i=2\mapsto 1-x}(y_{i}^{\prime}-y_{i}^{\prime \prime})(v_{x}-<v>)>0$. The first pair of expressions read as if the female allele population vector $Y$ is such that a mixed strategy $X^{\prime }$ would receive a higher  payoff against $Y$ than would $X^{\prime \prime}$, then the approaching speed of the system is faster toward $X^{\prime}$ than toward $X^{\prime \prime}$, in other words the tendency to move toward $X^{\prime}$ is higher than toward $X^{\prime \prime}$. Similar type of significance can be drawn to the second pair of expressions associated with the female allele population.
We see that $\lim_{x \rightarrow 0}\frac{\dot{x}}{x}$ and $\lim_{y \rightarrow 0}\frac{\dot{y}}{y}$ exist and are finite, i.e., the system of equations (\ref{eq:5}), is  a regular selection dynamics; therefore it is straightforward to verify that the replicator dynamics (\ref{eq:5}) is aggregate monotonic. Also, the time derivatives of bilinear average fitness functions $\pi_{1}(X,Y)$ and $\pi_{2}(Y,X)$ along the solution path to the replicator dynamics (\ref{eq:5}) through any given state $(X,Y)$ can be written as:
\footnotesize
\begin{eqnarray*}
 \dot{\pi}_{1}(X,Y) &\geq & \sum x_{i} ( \pi_{1}(e^{i},Y)-\pi_{1}(X,Y))^{2}+ m \sum y_{i} ( \pi_{2}(e^{i},X)-\pi_{2}(Y,X))^{2}, \nonumber \\
 \dot{\pi}_{2}(Y,X) &\geq & \frac{1}{M}\sum x_{i} ( \pi_{1}(e^{i},Y)-\pi_{1}(X,Y))^{2}+ \sum y_{i} ( \pi_{2}(e^{i},X)-\pi_{2}(Y,X))^{2}
\end{eqnarray*}
\normalsize
where $m \leq \frac{\pi_{1}(X,e^{j})}{\pi_{2}(e^{j},X)} \leq M $ or $\frac{1}{M} \leq \frac{\pi_{2}(Y, e^{i})}{\pi_{1}(e^{i},Y)} \leq \frac{1}{m}$, $m$ and ${M}$ being lower and upper bounds respectively. That is $\dot{\pi}_{1}(X,Y) \geq 0$ and $\dot{\pi}_{2}(Y,X) \geq 0$ with equality if and only if $(X,Y)$ belongs to the equilibrium solution set (the mathematical proof of our assertion is recorded in \ref{C}: Supplemental materials). Consequently, Proposition \ref{p4.2} is valid for all bimatrix games. And the inclusion relation connected to NEset (Nash equilibrium set) is applicable in the aspect of Fishman \cite{Fishman2008}.

\begin{figure}[t!]
\makebox[\textwidth][c]{
\begin{minipage}{.4\linewidth}
\centering
First Column \\[10pt]
\includegraphics[width=1.2\linewidth]{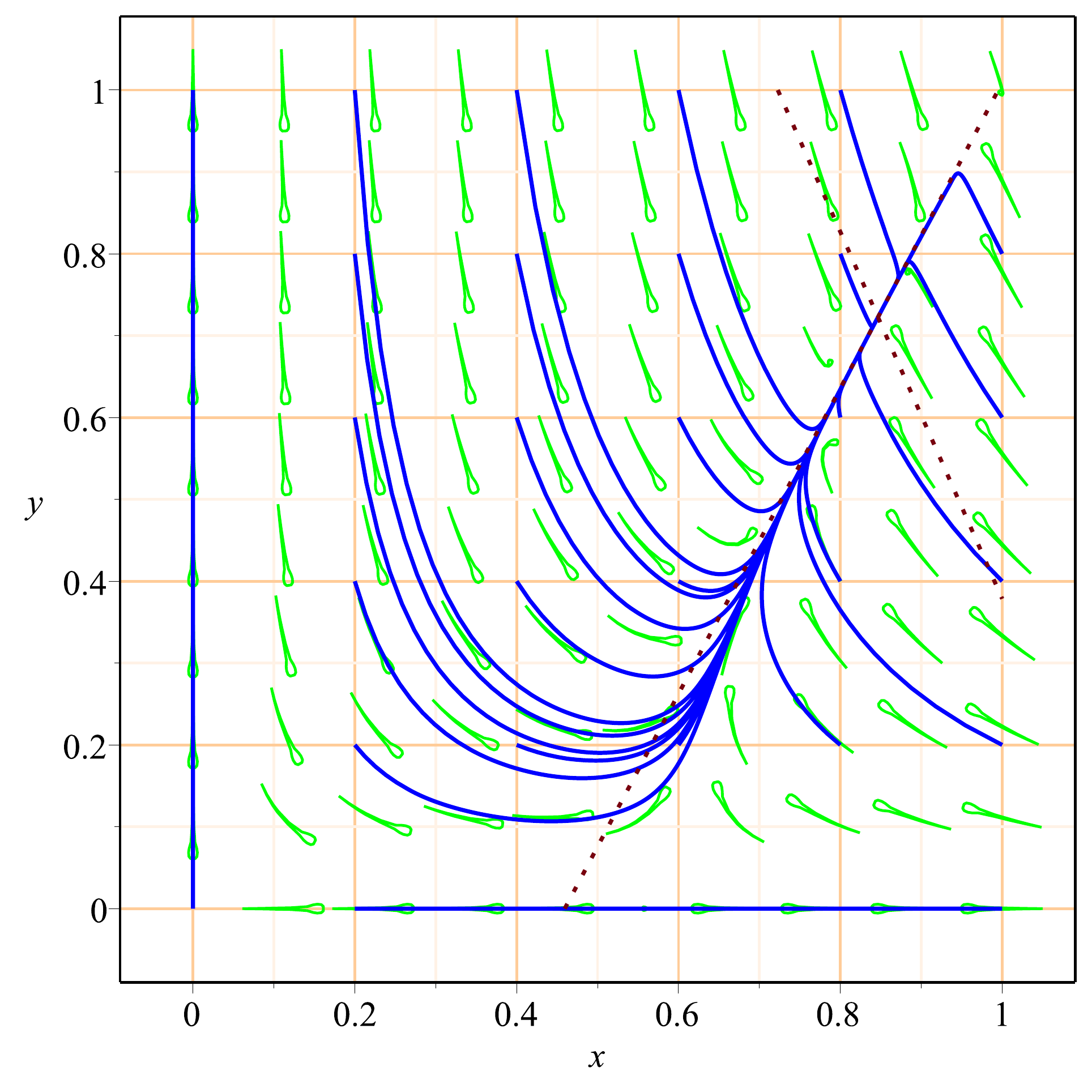}
\label{fig:test7}
\end{minipage}%
\hspace{1cm}
\begin{minipage}{.4\linewidth}
\centering
Second Column \\[10pt]
\includegraphics[width=1.2\linewidth]{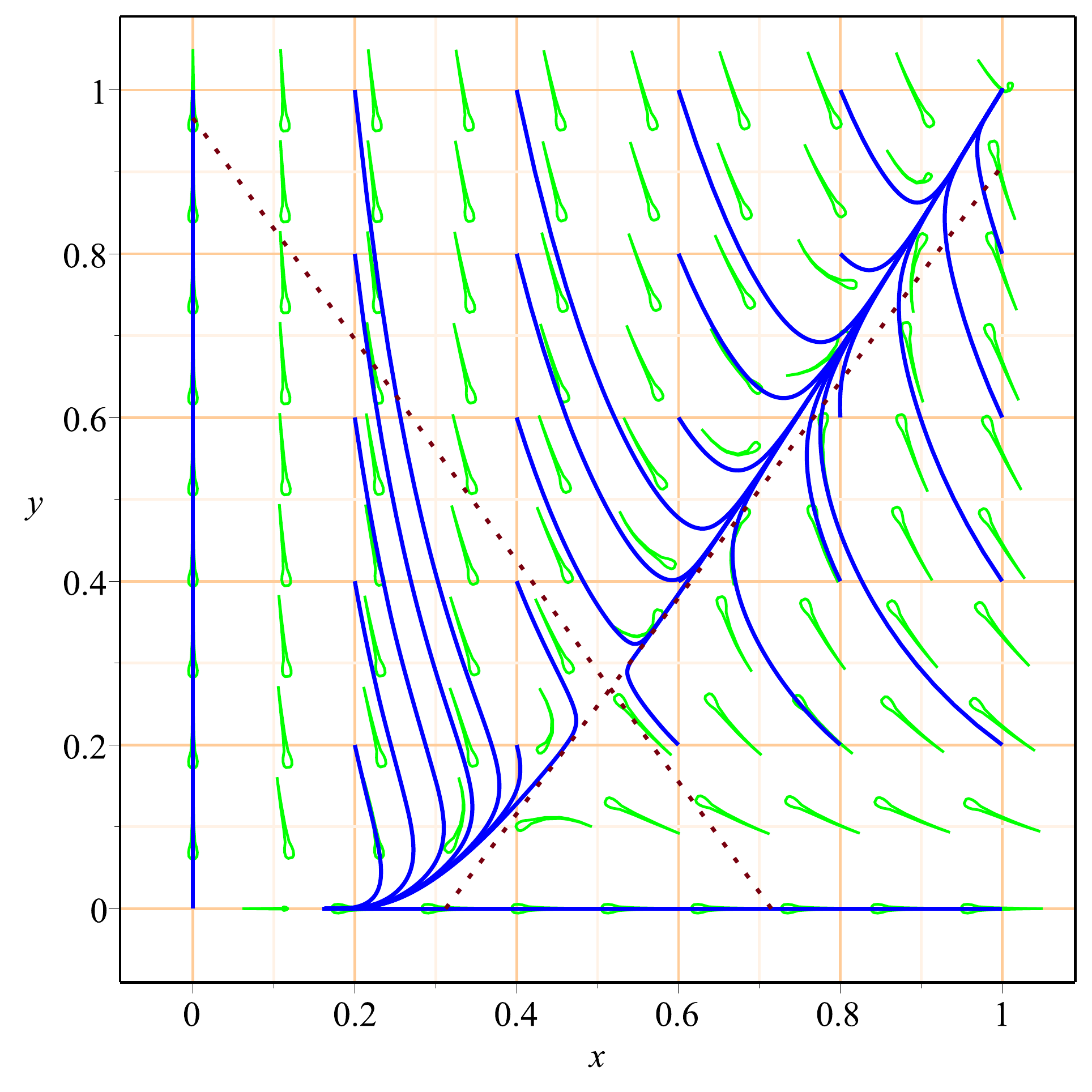}
\label{fig:test8}
\end{minipage}
\hspace{1cm}
\begin{minipage}{.4\linewidth}
\centering
Third Column \\[10pt]
\includegraphics[width=1.2\linewidth]{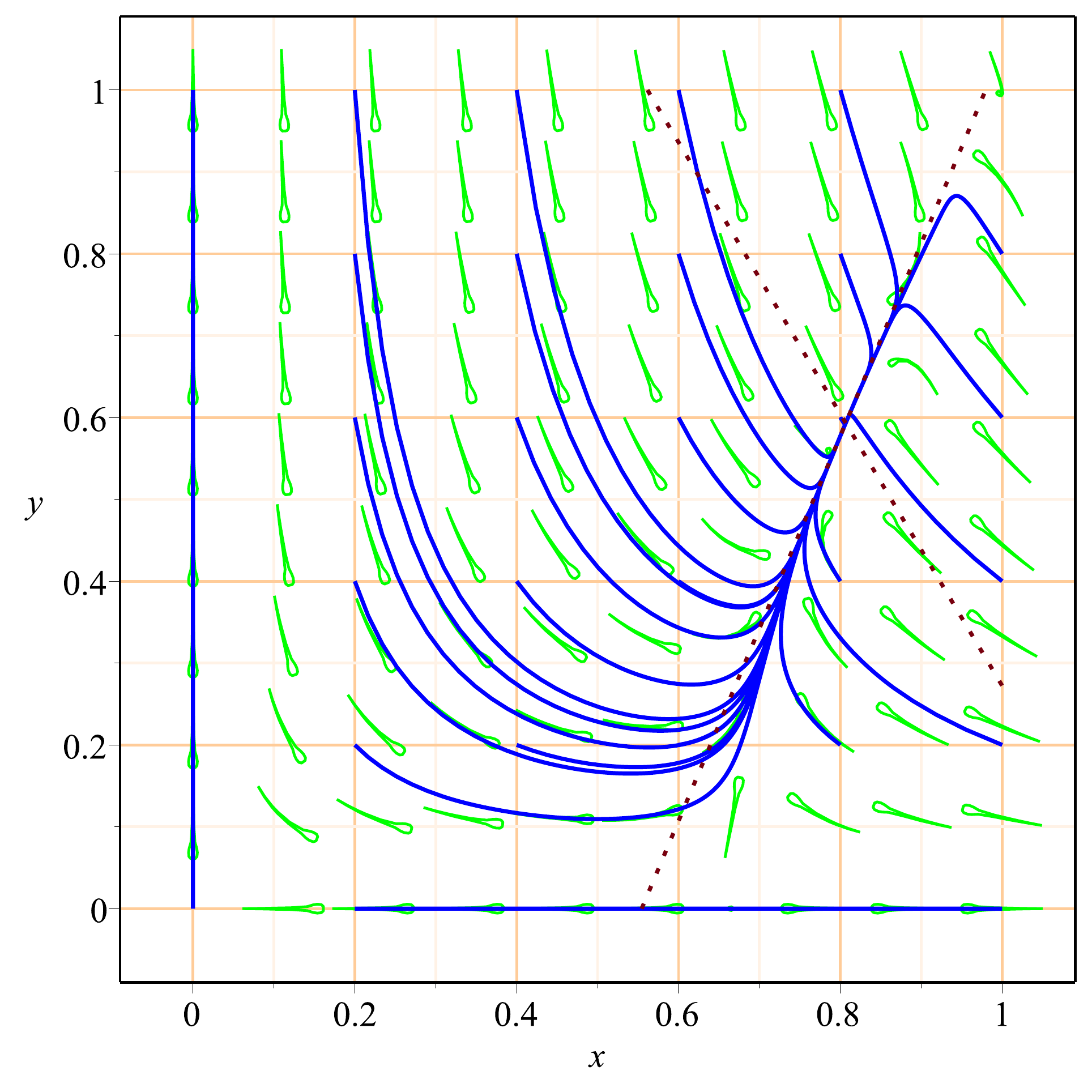}
\label{fig:test9}
\end{minipage}} \\
\makebox[\textwidth][c]{
\begin{minipage}{.4\linewidth}
\centering
\includegraphics[width=1.2\linewidth]{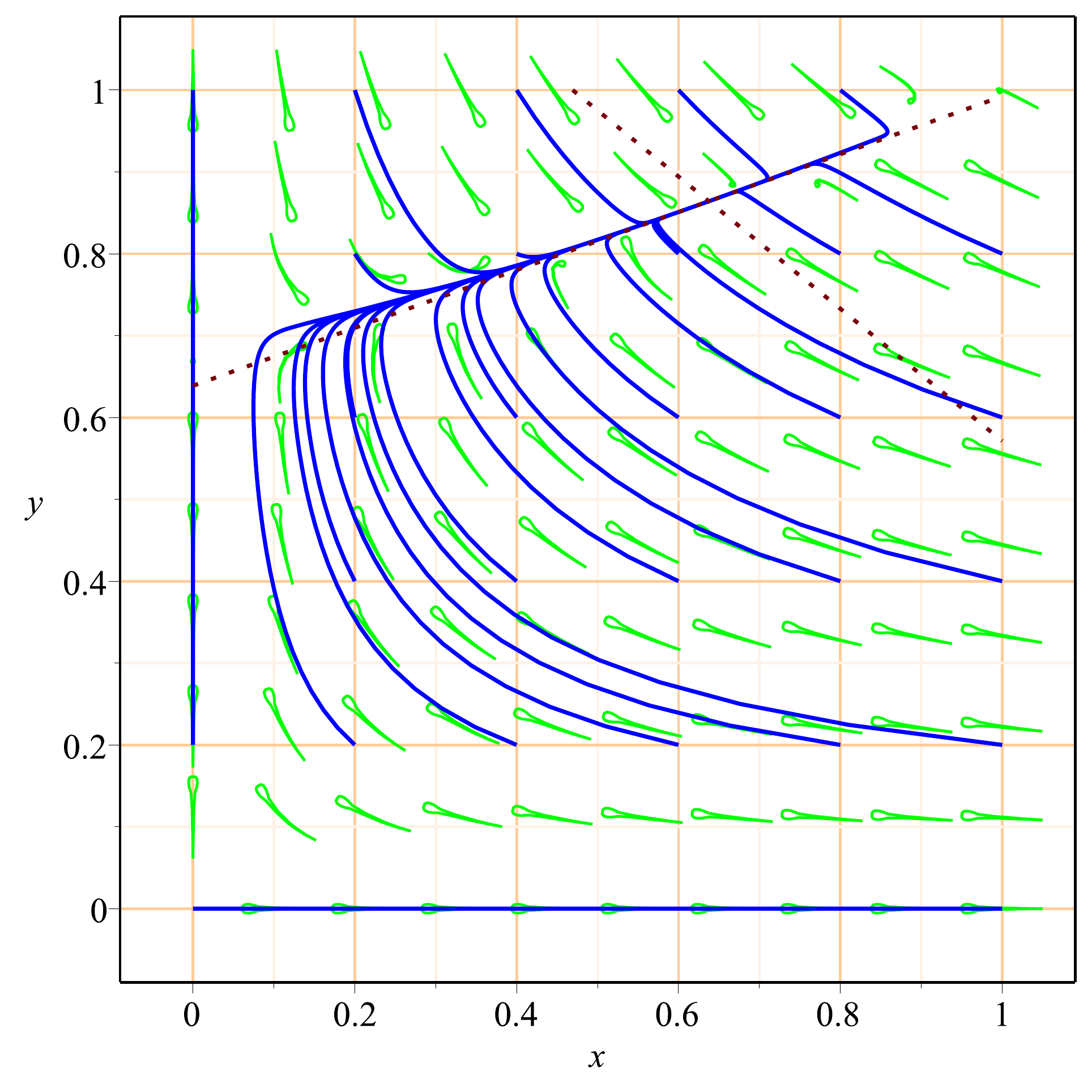}
\label{fig:test10}
\end{minipage}%
\hspace{1cm}
\begin{minipage}{.4\linewidth}
\centering
\includegraphics[width=1.2\linewidth]{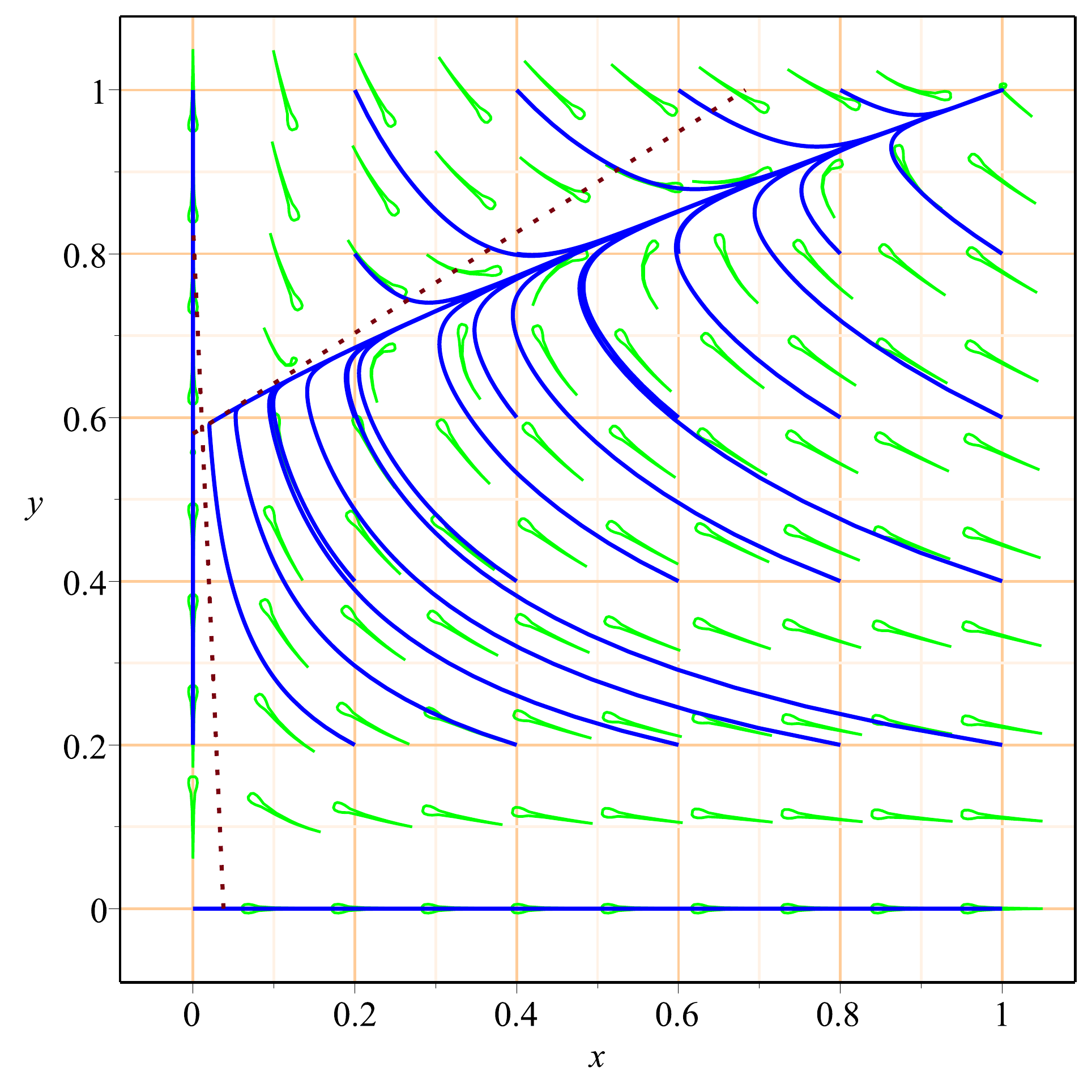}
\label{fig:test11}
\end{minipage}
\hspace{1cm}
\begin{minipage}{.4\linewidth}
\centering
\includegraphics[width=1.2\linewidth]{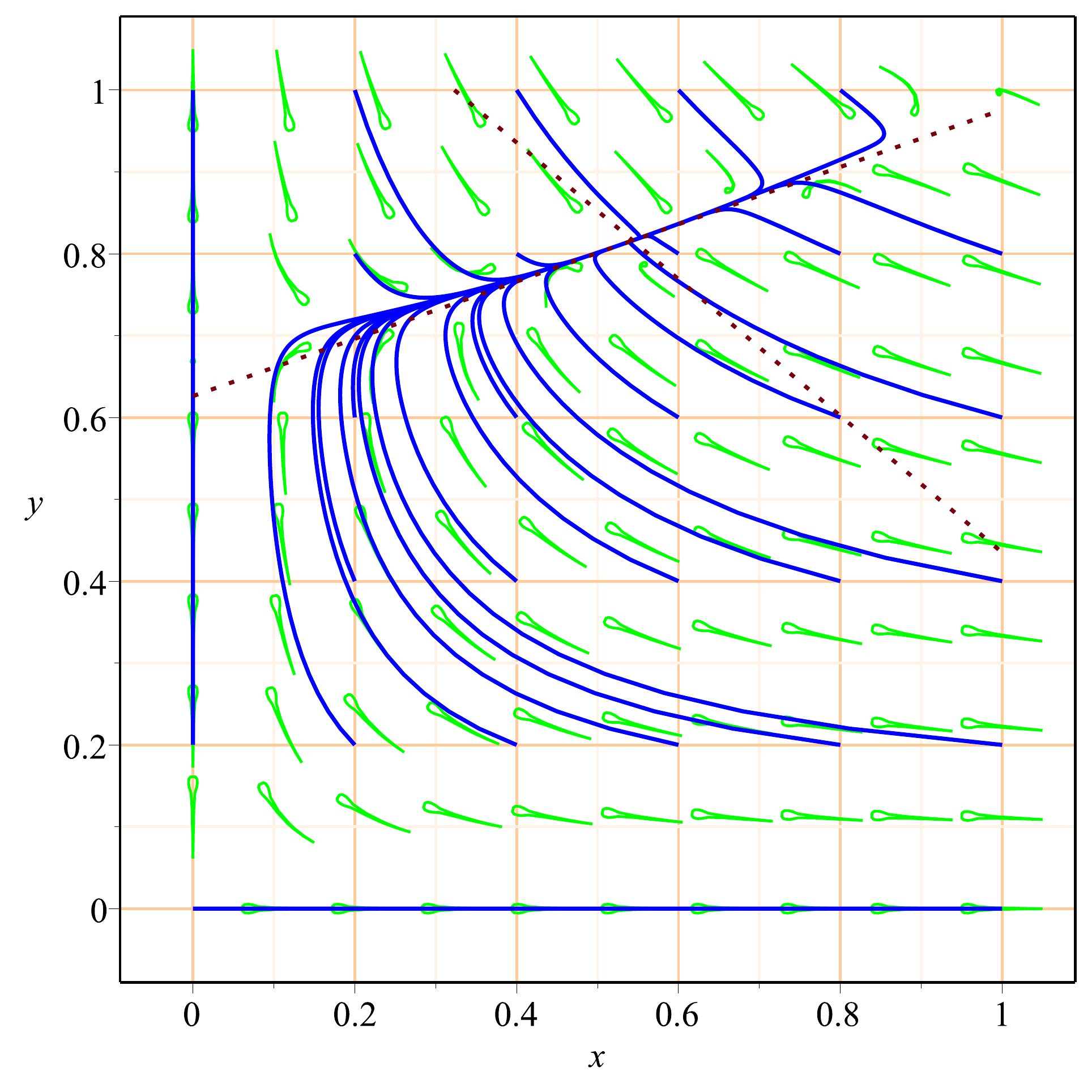}
\label{fig:test12}
\end{minipage}} 
\renewcommand{\figurename}{Fig.}
\caption[.]{\footnotesize {Array representation of segregation distortion effect on the nature and position of the internal equilibrium. Consider the initial array values of the fitnesses of Fig.\ref{fig:stability} with the same plot configuration; i.e., along the first, second and third columns the cases of dominant, recessive and overdominant selections have been depicted, respectively. In the array representation, the segregation configuration is carried out by the two sets of segregation ratio values $(k,\kappa)=(0.79, 0.05)$ and $(k,\kappa)=(0.05, 0.79)$ to consider along the first row and the second row respectively, where different alleles are favored in the two sexes. Although  $w_{21}$ and $v_{12}$ are undefined at any boundary point, some boundary points have been taken as initial points since the dynamical system is regular. The dotted lines represent the manifold (eigenspace). In recessive fitness, one of the manifolds is stable and other is unstable, whereas both manifolds are stable in the dominant and overdominant selections}.}\label{fig:Phase portrait1}
\end{figure}

We return to our main arena. In the fitness respect, favoring $A_{2}$ allele in the male population and $A_{1}$ allele in the female population, all the phase portraits in Fig.\ref{fig:Phase portrait1} have been plotted where the invading powers of $A_{1}$ male and $A_{1}$ female, however not linearly, depend on the factors $\mid k-w_{22} \mid$ and $\frac{1}{\mid \kappa+v_{11} \mid}$ respectively in dominant selection. In the overdominance, similar factors may be represented as $\mid k-\frac{1}{2}w_{12} \mid$ and $\mid \kappa-\frac{1}{2}v_{21} \mid$ in the male population and the female population respectively. At any set of segregation ratio values $(k,\kappa)$, other than Mendelian, the invading powers of $A_{1}$ male and $A_{1}$ female are both increased with increasing of the recessive fitness strength. It is a singularity against our expectation, and that can be explained by the high rates of invading power of $A_{1}$ female. On the other hand, at the Mendelian segregation cases, the frequency of $A_{1}$ female under the overdominance and dominance fitness environments is roughly affected by the fitness strength, and in contrast to that observation, instead of $A_{1}$ female, the frequency of $A_{1}$ male remains almost unchanged in the recessive environment.

Manifolds divide the whole state space into the four parts and size of each part is dependent on the set value of segregation ratios $(k, \kappa)$. It can be shown that at Mendelian segregation the sizes of the four parts are nearly equal to each other in most of the fitness environments. The observation expresses the fact that the evolution always tends to follow not only a $1:1$ sex ratio but also a $1:1$ different alleles ratio at particular gene locus. In most of the cases, if $k>\kappa$ and if the considered trajectories having the initial value ($i.v.$) condition, $y_{i.v.}>x_{i.v.}$, then those trajectories would have high valued radius of curvature; in other words, under this circumstance the frequency of $A_{1}$ female would vary in a wide range, while if $\kappa>k$ and if $x_{i.v.}>y_{i.v.}$ then opposite scenario would find -- the frequency of $A_{1}$ male would vary in a wide range ( see Fig.\ref{fig:Phase portrait1} ). All the observations having been pointed out in this and preceding paragraphs have strong evidences for validity in view of the numerical data analysis, but I have no mathematical proof (the details of these numerical evidences are provided in the \ref{C}: Supplemental materials).

Using calculus procedure to find out interior mixed Nash equilibrium points in a bimatrix game, we can here calculate a Nash equilibrium in mixed strategies, which is given by $((\frac{a_{6}}{a_{7}}, 1-\frac{a_{6}}{a_{7}}),(\frac{a_{1}}{a_{3}}, 1-\frac{a_{1}}{a_{3}}))$. The derivation is postponed to \ref{C}: Supplemental materials. At this point, the population mean fitness in both sexes is maximized over those variables that the alleles control. However, except the recessive fitness, in  the dominance and overdominance, it is not possible to determine such analytical Nash equilibrium because in those fitness structures $a_{1}$ and $a_{6}$ are always negative. The mentioning point explains the reality that $E_{5}$ is asymptotically stable in the dominance  and overdominance because there is no other Nash equilibrium. In whole feasible state space $(1,0)\times(0,1)$, at the point of numeral Nash equilibrium $E_{5}$, the population mean fitness in both  sexes is locally maximized, while due to the presence of analytical Nash equilibrium point, the instability of  $E_{5}$ in the recessive fitness structure  can be explained by the concept of the Fundamental Theorem of Natural Selection. Here, it should be kept in mind that, the space of segregation ratio values, $(k,\kappa)$, with the particular fitness values may yield an unstable polymorphic equilibrium which is not to be a Nash equilibrium.

From the different angle of evolutionary game theory, one can point out the three possibilities for the selection dynamics between two alleles $A_{1}$ and $A_{2}$ (strategies) to study the nature of fitness structure based on the behavior of polymorphic equilibrium $E_{5}$. The cases related to the three possibilities may be discussed in the following way \citep{Nowak2006, Hashimoto2014}:

\begin{itemize}
  \item $A_{1}$ and $A_{2}$ try to coexist in the female allele population while in the male allele population $A_{1}$ and $A_{2}$ try to reach bistable form. The observation corresponds to a saddle point $E_{5}$ in the recessive fitness structure ( see Fig.\ref{fig:Phase portrait1} ). And this is the case if $w_{11}>2(1-k)w_{12}r_{xy}$, $w_{22}>2kw_{12}$, $v_{11}<2(1-\kappa)v_{21}r_{yx}$ and $v_{22}<2\kappa v_{21}$. It is anticipated that the inequalities will be satisfied on the phase space trajectories along with for all points of the two manifolds. The outcomes of the selection dynamics depend on the initial condition.
  \item $A_{1}$ and $A_{2}$ coexist in both sexes, that corresponds to the stable $E_{5}$. And this is the case if $w_{11}<2(1-k)w_{12}r_{xy}$, $w_{22}<2kw_{12}$, $v_{11}<2(1-\kappa)v_{21}r_{yx}$ and $v_{22}<2\kappa v_{21}$. There is no doubt that paths of all trajectories on the state space $(1, 0)\times (0, 1)$ will follow the rule that governs simultaneously by four inequalities.
  \item The neutral case can be considered if the payoff bimatrices of both sexes are identical with identical rows and if the frequencies of alleles are same in both sexes where evolution dynamics do not move and every point of the interval $[0, 1]$ is an equilibrium point. However, it does not have any biological significance.
\end{itemize}

Finally, we complete this subsection to claim the Proposition \ref{p4.4}, that is true in accordance with the proof - outline of Gasull and Giacomini \cite{Gasull2013}.
\begin{pro}
The system of equations (\ref{eq:5}), has no limit cycle.\label{p4.4}
\end{pro}
\begin{proof}
To prove the result, first we write the system of equations (\ref{eq:5}), in the following form,
\begin{eqnarray*}
\dot{x}=x(g_{0}(x)+g_{1}(x)y), \hspace{5mm}
\dot{y}=y(h_{0}(x)+h_{1}(x)y)
\end{eqnarray*}
where $g_{0}(x)=(a_{1}x-a_{4})$, $g_{1}(x)=(a_{2}-a_{3}x)$, $h_{0}(x)=(a_{5}x-a_{8})$ and $h_{1}(x)=(a_{6}-a_{7}x)$. Now, if there exists a limit cycle, then $ g_{1}(x)\neq 0$, otherwise either $x=\bar{x}$ is an invariant line while $g_{0}(\bar{x})=0$ as well as $g_{1}(\bar{x})=0$ or we get a non-periodic parametric expression of x - component. That is, we can always assume that $g_{1}(x)$ does not vanish in the region where can have a periodic orbit.

Next, we consider the family of Dulac function $\psi(x,y)=y^{\lambda-1}Z(x)$, where $Z(x)$ is an unknown function and $\lambda$ is to be defined later. Thus, we have 
\begin{eqnarray*}
M(x,y)&=&\frac{\partial}{\partial x}[\psi(x,y) x(g_{0}(x)+g_{1}(x)y)]+\frac{\partial}{\partial y}[\psi(x,y) y(h_{0}(x)+h_{1}(x)y)]\\
      &=&[(xg_{0}(x)Z(x))^{'}+\lambda h_{0}(x)Z(x)\\ 
& & +((xg_{1}(x)Z(x))^{'}+(\lambda+1) h_{1}(x)Z(x))y]y^{\lambda-1}.
\end{eqnarray*}
The solutions to the differential equation 
\begin{eqnarray*}
(xg_{1}(x)Z(x))^{'}+(\lambda+1) h_{1}(x)Z(x)=0
\end{eqnarray*}
are 
\[ Z_{x_{0}}(x)=\frac{exp[-(\lambda+1)\int_{x_{0}}^x \frac{h_{1}(s)}{sg_{1}(s)} ds]}{xg_{1}(x)}\]
where $x_{0}>0$ is an arbitrary initial count. Therefore, taking the Dulac function $\tilde{\psi}(x,y)=y^{\lambda-1}Z_{x_{0}}(x)$ where $Z_{x_{0}}(x)$ is the solution to the above first-order differential equation, we can obtain after some simple algebraic manipulation that
\begin{eqnarray*}
M(x,y)&=&\frac{Z_{x_{0}}(x)}{g_{1}(x)}\Bigl[xg_{1}^{2}(x)\Bigl(\frac{g_{0}(x)}{g_{1}(x)}\Bigr)^{'}+\lambda( g_{1}(x)h_{0}(x)-g_{0}(x)h_{1}(x))\\
&  &  -g_{0}(x)h_{1}(x)\Bigr].
\end{eqnarray*}
Thus, we can choose the value of $\lambda$ such that $M(x,y)$ does not change sign in $(1,0)\times (0, 1)$ region, since all the involved  functions are bounded in the considering region. Hence, by Bendixson-Dulac Theorem, the claim is proved.
\end{proof}

\subsection{Population genetics in group selection framework}

We start the subsection with some redefining cornerstones of game theory in the flavor of group structure. The redefining concept is based on the analytical thinking of  the group selection of Jensen and Rigos \cite{Jensen2012}, with adopting the two acronyms: $NEGS$ and $ESSGS$.

As at the Nash equilibrium average fitness is optimized and it strengthens the equilibrium concept, therefore $(p^{\ast},q^{\ast})$ is a Nash equilibrium with group selection ($NEGS$) if
\begin{eqnarray*}
(p^{\ast},1-p^{\ast}).(\pi_{C}(p^{\ast},q^{\ast}), \pi_{D}(p^{\ast},q^{\ast}))\geq (p,1-p).(\pi_{C}(p^{\ast},q^{\ast}), \pi_{D}(p^{\ast},q^{\ast})), \nonumber \\
(q^{\ast},1-q^{\ast}).(\pi_{C}^{'}(q^{\ast},p^{\ast}), \pi_{D}^{'}(q^{\ast},p^{\ast}))\geq (q,1-q).(\pi_{C}^{'}(q^{\ast},p^{\ast}), \pi_{D}^{'}(q^{\ast},p^{\ast})) \hspace{1.5mm} 
\end{eqnarray*}
for all $(p,q)$. The $(p^{\ast},q^{\ast})$ is a best reply to itself in the both populations, and if it is unique then the Nash equilibrium is called strict. On the other hand, we have already noted in the preceding subsection that the most commonly used solution concept in evolutionary game theory is that of an evolutionarily stable strategy ($ESS$) and the central idea behind the $ESS$ concept is that of non-invasion. Here, $ESS$ which is usually defined in games with random mating/matching can be to include non-random mating in the following way: A strategy $(\tilde{p},\tilde{q})$ is an evolutionarily stable strategy with group selection ($ESSGS$), if for each $(p,q)$, not equal to $(\tilde{p},\tilde{q})$, there exists $\tilde{\epsilon}_{(\tilde{p},\tilde{q})}$ such that
\begin{eqnarray*}
(\tilde{p},1-\tilde{p}).(\pi_{C}(\mathscr{P},\mathscr{Q}),\pi_{D}(\mathscr{P},\mathscr{Q}))>(p,1-p).(\pi_{C}(\mathscr{P},\mathscr{Q}), 
\pi_{D}(\mathscr{P},\mathscr{Q})), \\
(\tilde{q},1-\tilde{q}).(\pi_{C}^{'}(\mathscr{Q},\mathscr{P}), \pi_{D}^{'}(\mathscr{Q},\mathscr{P}))>(q,1-q).(\pi_{C}^{'}(\mathscr{Q},\mathscr{P}), \pi_{D}^{'}(\mathscr{Q},\mathscr{P})) \hspace{1.5mm} 
\end{eqnarray*}
for all $\epsilon \in (0,\tilde{\epsilon}_{(\tilde{p},\tilde{q})})$, where $\mathscr{P}=\epsilon p+ (1-\epsilon)\tilde{p}$ and $\mathscr{Q}=\epsilon q+ (1-\epsilon)\tilde{q}$. The algebraic inequalities express that with new population composed of $(1-\epsilon) (\tilde{p},\tilde{q})$-strategists and $\epsilon (p,q)$-strategists, a small but measurable (of measurable $\epsilon$ up to $\tilde{\epsilon}_{(\tilde{p},\tilde{q})}$) group of individuals using any strategy $(p,q)$, exclusion of $ESSGS$ $(\tilde{p},\tilde{q})$, can not invade the male population and the female population where all individuals use an $ESSGS$ $(\tilde{p},\tilde{q})$. Individuals using $ESSGS$ will get a higher expected payoff than the invaders. And, hence, with similar fashion we can define Local Superiority: $(\bar{p},\bar{q})$ is called locally superior if there exists a neighborhood $U$ of $(\bar{p},\bar{q})$ such that exclusion of $(\bar{p},\bar{q})$  for all $(p,q)$,
\begin{eqnarray*}
(\bar{p},1-\bar{p}).(\pi_{C}(p,q),\pi_{D}(p,q))>(p,1-p).(\pi_{C}(p,q), 
\pi_{D}(p,q)), \\
(\bar{q},1-\bar{q}).(\pi_{C}^{'}(q,p), \pi_{D}^{'}(q,p))>(q,1-q).(\pi_{C}^{'}(q,p), \pi_{D}^{'}(q,p)). 
\end{eqnarray*}
Clearly,  $(\tilde{p},\tilde{q})$  is an $ESSGS$ if and only if it is locally superior.

\begin{figure}[b!]
\makebox[\textwidth][c]{
\begin{minipage}{.4\linewidth}
\centering
\includegraphics[width=1.2\linewidth]{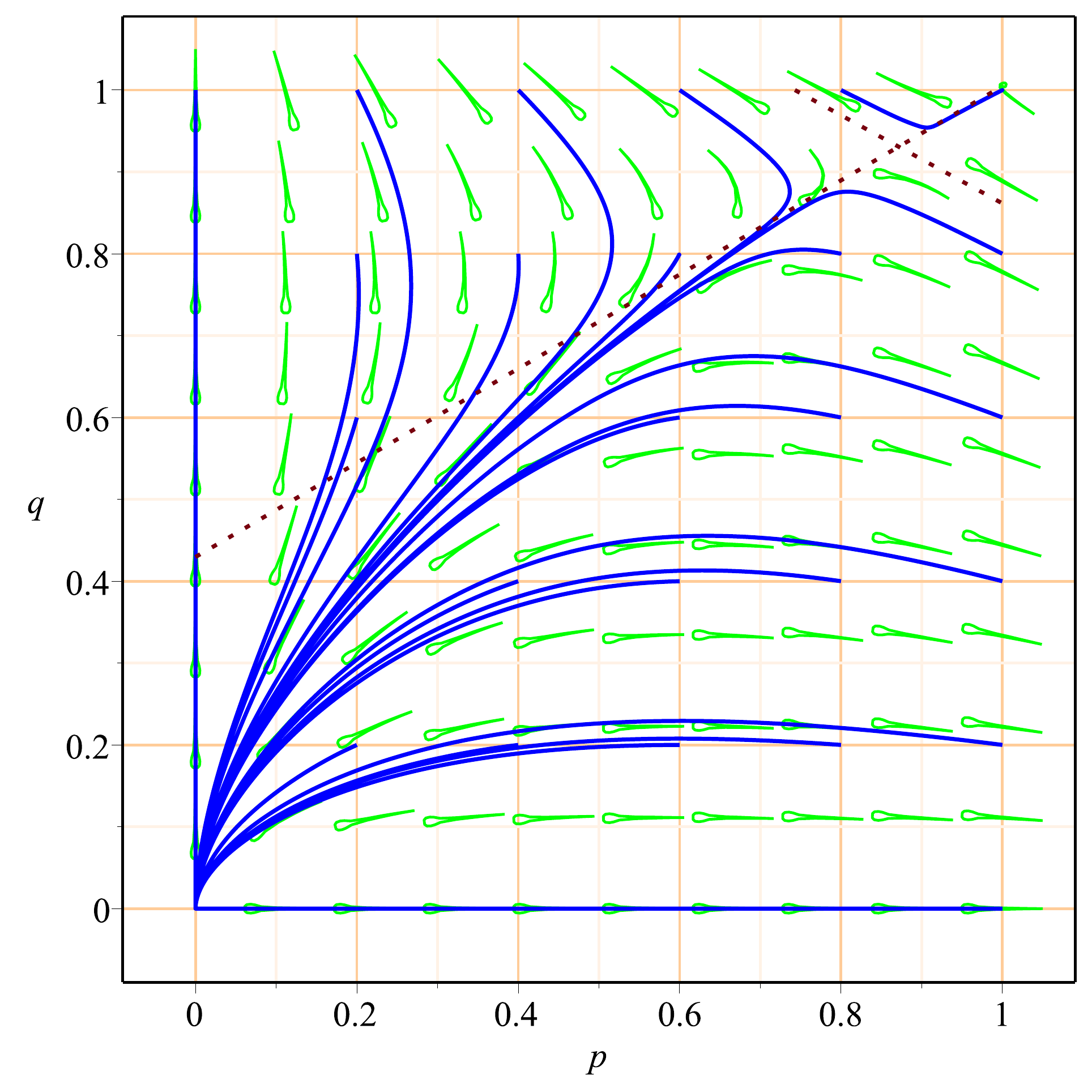}
\label{fig:test104}
\end{minipage}%
\hspace{1cm}
\begin{minipage}{.4\linewidth}
\centering
\includegraphics[width=1.2\linewidth]{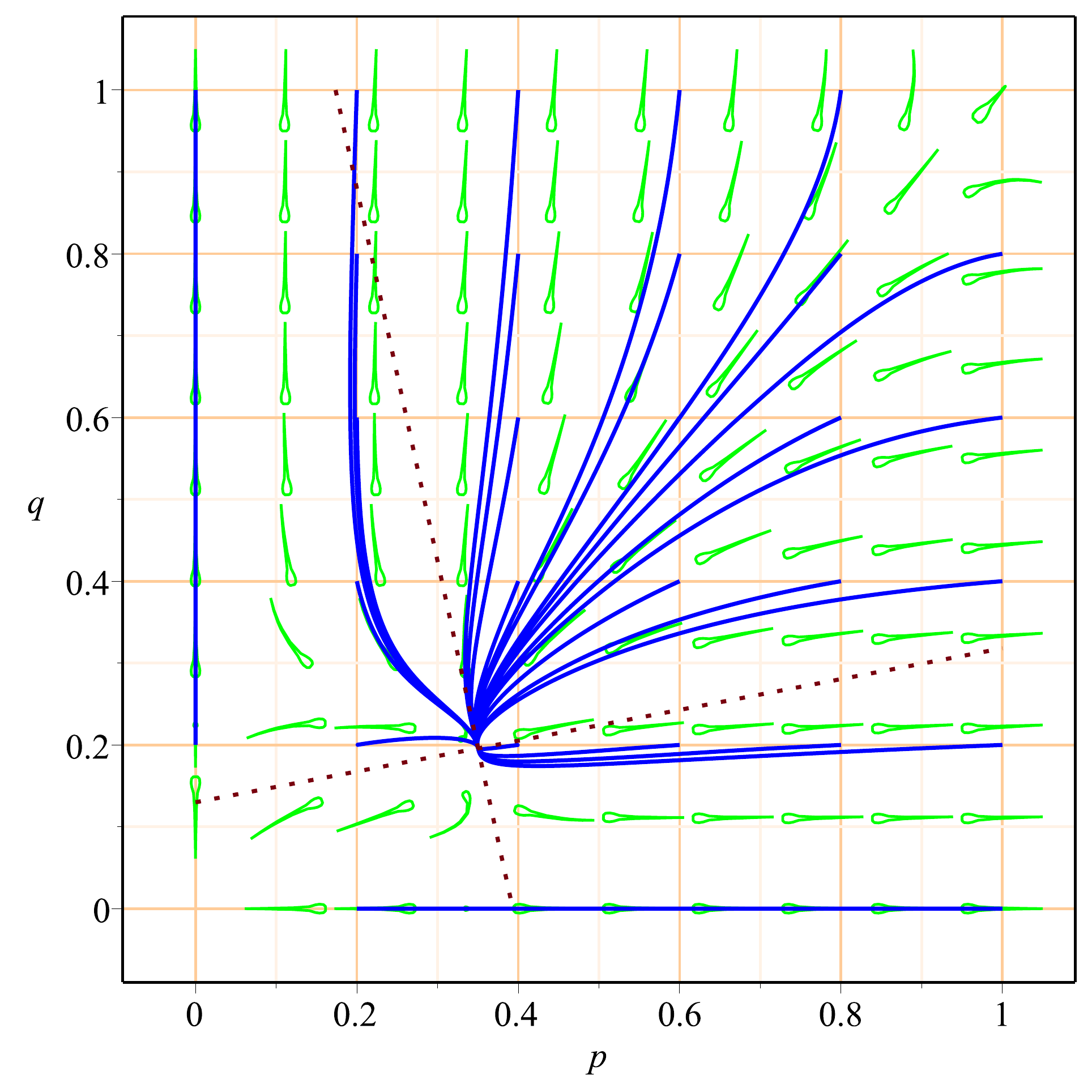}
\label{fig:test105}
\end{minipage}
\hspace{1cm}
\begin{minipage}{.4\linewidth}
\centering
\includegraphics[width=1.2\linewidth]{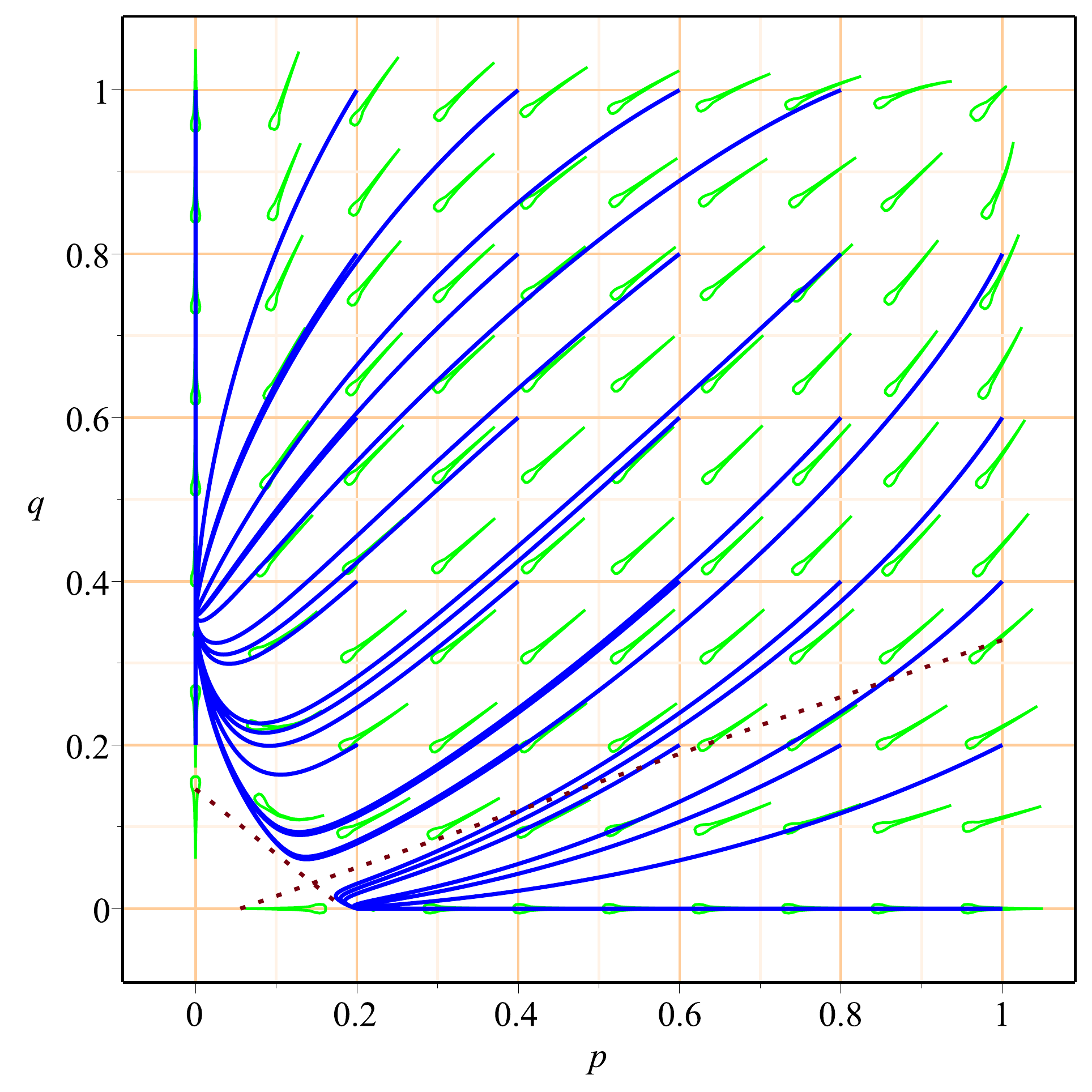}
\label{fig:test106}
\end{minipage}} 
\renewcommand{\figurename}{Fig.}
\caption[.]{\footnotesize {Row representation of segregation distortion effect on the nature and position of the internal equilibrium. The same plot configuration of  viability effects of Fig. \ref{fig:GSstability} has been followed, where in the first, second and third figures, the sets of segregation ratio values are  $(k, \kappa)=(0.8, 0.95)$, $(k, \kappa)=(0.79, 0.75)$ and $(k, \kappa)=(0.90, 0.92)$, respectively}.}\label{fig:Phase portrait2}
\end{figure}
We now come to the phase portrait structure. Specifically, an equilibrium solution $(p(t),q(t))=E_{6}$ to equations (\ref{eq:6}) is asymptotically stable if every solution to equations (\ref{eq:6}) which starts sufficiently close to $E_{6}$ at initial time, $t=0$, not only remains close to $E_{6}$ for all future time but ultimately approaches $E_{6}$ as $t$ approaches infinity. More precisely, $E_{6}$ is asymptotically stable if there exists a neighborhood $U$ of $E_{6}$ such that
\( (p(0),q(0))\in U \Rightarrow \lim_{t \rightarrow  \infty}(p(t),q(t))=E_{6} \). Therefore, the conclusion can be taken as the following proposition:

\begin{pro}
If the equilibrium solution $E_{6}$ of equations (\ref{eq:6}) is asymptotically stable, then it is locally superior as well as $ESSGS$. 
\end{pro}
As, the system of equations (\ref{eq:6}), is a regular monotonic selection dynamics, $E_{6}$ is a $NEGS$, and as $E_{6}$ is not solution to the system of equations: \( \frac{\partial \bar{\pi}(p,q)}{\partial p}=0 \hspace{1mm} \mbox{and} \hspace{1mm} \frac{\partial \bar{\pi}^{'}(q,p)}{\partial q}=0 \), i.e., $NEGSset$ is not singleton, we obtain the following chain of inclusions:
\[ ESSGSset \subseteq \mbox{asymtotically stable set} \subseteq NEGSset \subseteq \mbox{equilibrium solution set} \]

Hence, if  it is imposed that at the boundary, all the involved biological parameters are defined, then we observe that in the recessive selection environment of non-random mating, the equilibrium point $E_{0}$ corresponds to $ESSGS$ while $E_{6}$ is a saddle point (see Fig.\ref{fig:Phase portrait2}). The size of the basin of attraction of $E_{0}$ is larger than the effective region of $E_{6}$, where the position of $E_{6}$ is in the vicinity of $E_{3}$ and with high set values of $(k,\kappa)$, $E_{6}$ approaches to $E_{3}$; in other words, the whole state approaches to $A_{2}$ allele-state. Below the bifurcation value of mating parameter $\alpha$, in the overdominant selection, $E_{6}$ turns to $ESSGS$ -- alleles $A_{1}$ and $A_{2}$ are coexisted in both the male and female populations -- whereas the corner points are unstable and $E_{6}$ tends to $E_{0}$ with high values of $(k, \kappa)$ that being confined in the feasible region. Within the unstable feasible region of $(k, \kappa)$ in the overdominant selection, $E_{6}$ becomes a saddle point, placed near about $E_{0}$. It is seen that the two boundary points on $p$-axis and $q$-axis act as $ESSGS$ where the point on $q$-axis has a larger basin of attraction than the other point. Here is no doubt that at Mendelian segregation, the possibility to find out a polymorphic equilibrium point is to be relatively low. As modeling results are obtained mainly based on the varying values of two segregation ratio parameters and one mating parameter under the assumption of the Fundamental Theorem of Natural Selection those have strong biological influences on the stability of genotype or phenotype character, thus every single result has its own biological relevance which is intuitively understandable.
\nopagebreak
\section{Discussion and future work}
Generally, population genetics deals with discrete generations where the population moves into Hardy-Weinberg proportions in one generation in such a way that parents do not participate in further reproduction once the daughter generation is formed and are on longer counted as part of the population. This is a reasonable assumption only for some real-world populations in which organisms breed synchronously and only once in their lifetime. However, in the case of reproduction and / or overlapping generations, to develop the continuous genetic realistic model is quite significant, by focusing the assumption that births and deaths can take place at any instant. Referred to this many continuous models have been formulated in respect of various mathematical formulations (see Ewens \citep{Ewens2004}, Felsenstein \citep{Felsenstein2011} and references therein). Among them, the mathematical structure of replicator selection dynamics is most realistic one, which has been shown in this article by revealing the fact that, without imposing any additional constraint, Hardy-Weinberg frequencies hold in replicator model in the sense of, $\dot{\pi}_{1}(X,Y)=0$ and $\dot{\pi}_{2}(Y,X)=0$ if  and only if the population in equilibrium. Side by side, it is also noticed that the evolution occurs fastest when the variance in fitness is maximized.

In the present article, a one-locus  model of continuous replicator selection dynamics deriving from the discrete genetic model of sex-specific viabilities under the consideration of nonequivalence of the viabilities of reciprocal heterozygotes that might be observed at an imprinted locus is studied in aspect of the sex-specific meiotic drive. The study was motivated to construct the set of link-results between population genetics and evolutionary game theory where the group selection theory is also incorporated. It is known that in the discrete genetic model, in the case of Mendelian segregation, the change in gene frequency is linearly related to the rate of change of population mean viability fitness \citep{Ubeda2004}; because of that the stable polymorphic equilibrium maximizes the population mean fitness. Analytically, in the game theory standpoint, it is not possible to show that under the feasible values of biological parameters, there exists a Nash equilibrium point or a point of $ESS$, the point at which the population mean fitness is maximized, over those variables that the allele controls. Only the special case of underdominance we find out the point of pure $ESS$. However, regarding the segregation in both scenarios of dominant and overdominant selections, $E_{5}$ corresponds to the point of $ESS$ as the point is asymptotically stable in the dynamical structures of those selections where the Fundamental Theorem of Natural Selection is held. That is, the model suggests that in asymmetric games mixed $ESS$ can exist. It simply supported the conclusion of Binmore and Samuelson \cite{Binmore2001} of the existence of a mixed $ESS$ in an asymmetric game in contradiction to the classic theorem of Selten \cite{Selten1980}.

Quantitative genetics has an inherited pattern of thought to construct genetic models of the evolutionary process in flavor of evolutionary game dynamics. In this branch, construction of models is initiated from the phenotype levels, and selection terms are decomposed into the components to relate phenotypes to fitness (selection differentials) and genes to phenotypes (heritability). Queller \cite{Queller1992} provides a general rule for determining when such decomposition is justified and shows how Price's covariance equation is associated with standard quantitative genetic results and derives quantitative genetic equations for inclusive fitness and group selection. Price's equation represents the dynamics of changing gene frequencies, and inclusive fitness theory provides conditions for the evolutionary success of a gene. Hamilton's inclusive fitness theory says that if an allele in individual $\mathscr{A}$ increases the fitness of individual $\mathscr{B}$ whose degree of relatedness to $\mathscr{A}$ is $r$ and if $c$ is the cost to the altruist and $b$ is the benefit to the recipient of the altruistic behavior, then the allele will be favored by natural selection if $br>c$ where $r$ is generally defined as the probability that the recipient of altruistic behavior has a copy of focal allele. As Hamilton's rule presupposes weak selection (i.e., population gene frequencies do not vary appreciably in a single reproduction period), successful allele related to inclusive fitness will not move to fixation in the genome. Moreover, mutation can act in such a way that the frequency of focal allele remains unchanged. Although the inclusive fitness concept is more general than kin selection, in the majority of the cases, it has been shown that Hamilton's rule in principle has a relationship with genealogy or kin selection. Nowak et al. \cite{Nowak2010a} point out that cooperators are favored over defectors for weak selection if a condition holds that is of the form: $something>\frac{c}{b}$ which is nothing but the mathematical expression of inclusive fitness, but typically that $something$ is not always related to relatedness. In this present paper, the inclusive fitness effect is neglected in comparison with the viability effect and the mating parameter $\alpha$ of population structure, a specific form  of that $something$ other than relatedness, is dependent on the social structure of the reproduction population (see Gintis \citep{Gintis2014} and references therein).

Since the population structure is a function of allele frequencies, its mathematical formalization implies group selection procedure of alleles. That is, in case of the non-random mating, the single-locus-multiple-allele models will yield more than one polymorphic equilibrium (see Gokhale and Traulsen \citep{Gokhale2012}) whereas only one polymorphic equilibrium, $E_{6}$ exists in the single-locus-two-allele model. Multiple allele configurations are exceedingly complex where alleles are usually sequences that code for genes. Thus, to find out the realistic population structure as a function of allele sequences would be a topic for future research work along with to investigate how the number of alleles (size of groups) influences the polymorphic equilibria. 

Feldman and Otto \cite{Feldman1991} presented the dynamical structure of one- and two-locus multiple-allele viability system where in consideration of the role of recombination the sex-determination system in Aedes aegypti was modeled in terms of two linked genes and the treatment of segregation distorter complex had also included at second locus. On the other hand, neglecting the important factor of recombination Eshel \cite{Eshel1985} claims that the selection on unlinked modifiers favors Mendelian segregation at a polymorphic locus. Allowing sex-specific segregation distortion, Eshel's model is further generalized by \'{U}beda and Haig \cite{Ubeda2005}, but the generalized model shows that natural selection favors departure from Mendelian expectations. That is, the instability of Mendelian segregation requires a new view for a satisfactory explanation of why Mendelian segregation is the rule. At this stage, it is natural, the question then arises as to could the modified replicator selection dynamics by adding recombination (see Gaunersdorfer et al. \citep{ Gaunersdorfer1991}) explain the Mendelian-segregation's ubiquity? In order to set up new perspectives between population genetics and evolutionary game theory, answering this conundrum would be a challenging work.

\section*{Acknowledgments}
I would like to thank one of the anonymous reviewers and the handling editor for valuable comments and constructive suggestions on the standard of the presentation and explanation of the manuscript.
\appendix
\section{Replicator dynamics in group selection framework}\label{A}
To implement the group construction view of van Veelen \cite{Veelen2011} in a formulation procedure of the group selection dynamics between a male allele population and a female allele population, we consider that two sexes of $m$ and $n$ alleles form separately two groups of alleles of size $(m-1)+(n-1)$  those adopt two strategies (and  interpret phenotypic characters) --  $C$, defines the character of cooperating and $D$, links to the character of defecting; i.e., each group in each sex,  consists of $(m-1)$ alleles of male and $(n-1)$ alleles of female \footnotemark[1], and after the group formation, we assume that all alleles (entities) of the group, irrespective of their origin, are the alleles of the sex which forms the group. \footnotetext[1]{Wikipedia (\url{http://en.wikipedia.org/wiki/Allele}, Proposed since June 2014): \em{A population or species of organisms typically includes multiple alleles at each locus among various individuals.....For example, at the gene locus for the ABO blood type carbohydrate antigens in humans.....It is now known that each of the A, B, and O alleles is actually a class of multiple alleles with different DNA sequences that produce proteins with identical properties: more than $70$ alleles are known at the ABO locus.}}For each sex, groups are of any composition and they are formed by a non-random way in respect of the two strategies. In the course of the group formation by every allele, we use indexing of alleles in the both sexes. The $m$ alleles in the male are indexed as $0,1,2,....,m-1$ and similarly to $m$ alleles, for the $n$ alleles of the female, we give indices of $0,1,2,....,n-1$. As genotype is an allelic pair, at a single locus diploid population the numerical value of $m$ will necessarily be equal to $n$; and the necessity also explains the fact why every allele of both sexes does not participate in all the group formations at a time. Intention of introducing the two distinct counters is to represent the mixing groups of alleles of the sexes. Here, the group selection dynamics proceeds through the two steps at each time division: in the first step, the distinct and identical groups are formed by alleles at the particular gene locus of the both sexes with assortative copy, and, in the second step, after the group formation, strategists are selected by the viability effect where the effect is influenced by the group frequencies. To keep things as simple as possible,  realizable other biological influenced factors are ignored in the considering framework.  

Frequencies of the different types of groups are denoted by $f_{i,j}$ in the male allele population and by $\nu_{j,i}$ in the female allele population where $i=0,1,.....m-1, j=0,1,.....n-1$, and are defined as $f_{i,j}$ and $\nu_{j,i}$ being the frequency of groups  with $(i+j)$ $C$ alleles and $m+n-(i+j+2)$ $D$ alleles in it. Subscript $i$ denotes the number of male-$C$ alleles and $j$ denotes the number of female-$C$ alleles in the groups. As population states are to be characterized by the group frequencies, the frequencies have to satisfy the following conditions: $0\leq f_{i,j} \leq 1 \mbox{ for all}$ $i,j $ and \( \sum_{i=0}^{m-1} \sum_{j=0}^{n-1}f_{i,j}=1\) in male allele population, and $0\leq \nu_{j,i} \leq 1 \mbox{ for all}$ $ i,j $ and \( \sum_{j=0}^{n-1}\sum_{i=0}^{m-1} \nu_{j,i}=1\) in female allele population, where the marginal frequencies can be defined as $f_{i}=\sum_{j=0}^{n-1}f_{i,j}$  and $\nu_{j}=\sum_{i=0}^{m-1}\nu_{j,i}$. $f_{i,j}$ and $\nu_{j,i}$ are confined in the simplex $\Delta_{1}$ of $(m+n-1)$-dimension and the simplex $\Delta_{2}$ of $(m+n-1)$-dimension, respectively.

\sloppy
Hence, if the frequencies of the male-$C$ alleles and the  female-$C$ alleles are denoted by $p$ and $q$ respectively, then according to the group formation, the frequencies are calculated as 
\[ p=\frac{1}{m+n-2}\sum_{i=0}^{m-1}\sum_{j=0}^{n-1}(i+j)f_{i,j}  \mbox{ and } q=\frac{1}{m+n-2}\sum_{j=0}^{n-1}\sum_{i=0}^{m-1}(i+j)\nu_{j,i}.  \]  
Now, we let $\hat{f}$ and $\hat{\nu}$ be population structure functions on the simplexes  $\Delta_{1}$ and $\Delta_{2}$ respectively, and the corresponding mapping expressions are $\hat{f}:[0,1] \times [0,1]\rightarrow \Delta_{1}$ and $\hat{\nu}:[0,1]\times [0,1]\rightarrow \Delta_{2}$. By the following relations, the functions can be associated with frequencies $p$ and $q$, 
\[ \frac{1}{m+n-2}\sum_{i=0}^{m-1}\sum_{j=0}^{n-1}(i+j)\hat{f}_{i,j}(p,q)=p \hspace{1mm} \mbox{for all}\hspace{1mm} (p,q)\in [0,1]\times[0,1], \]
\[ \frac{1}{m+n-2}\sum_{j=0}^{n-1}\sum_{i=0}^{m-1}(i+j)\hat{\nu}_{j,i}(q,p)=q \hspace{1mm}\mbox{for all}\hspace{1mm} (q,p)\in [0,1]\times [0,1] \] 
where $\hat{f}_{i,j}(p,q)$ and $\hat{\nu}_{j,i}(q,p)$ are the components of a population structure function in males and a population structure function in females respectively; and, therefore, the sets of population structures in the male allele population and female allele population are obtained as 
\[S_{1}=\{\hat{f}:\rightarrow \Delta_{1} \mid \frac{1}{m+n-2}\sum_{i=0}^{m-1}\sum_{j=0}^{n-1}(i+j)\hat{f}_{i,j}(p,q)=p \hspace{1mm} \forall (p,q) \in [0,1]\times[0,1] \}, \]  
\[ S_{2}=\{\hat{\nu}:\rightarrow \Delta_{2} \mid \frac{1}{m+n-2}\sum_{j=0}^{n-1}\sum_{i=0}^{m-1}(i+j)\hat{\nu}_{j,i}(q,p)=q  \hspace{1mm} \forall (q,p) \in [0,1]\times [0,1] \}. \]

For males, the game payoffs (i.e., fitnesses) are denoted by $\pi_{C,i,j}$ and $\pi_{D,i,j}$ which are the payoffs to a $C$ allele and  a $D$ allele respectively when there are total (i+j) $C$ alleles in the group, and those for females are $\pi_{C,j,i}^{'}$ and $\pi_{D,j,i}^{'}$. 
Hence, under the population structures $\hat{f}$ and $\hat{\nu}$, the expected fitness payoffs of playing strategy $C$ and strategy $D$ in the male allele population and female allele population can be computed in a straightforward way:

 \[  \Bigl[\begin{array}{ll} \overline{\pi}_{C} \\ \overline{\pi}_{D} \\ \overline{\pi}_{C}^{'} \\ \overline{\pi}_{D}^{'} \end{array}\Bigr]= \Bigl[\begin{array}{ll} \frac{1}{(m+n-2)p}\sum_{i=0}^{m-1}\sum_{j=0}^{n-1}(i+j) \hat{f}_{i,j} (p,q) \pi_{C,i,j} \\ \frac{1}{(m+n-2)(1-p)}\sum_{i=0}^{m-1}\sum_{j=0}^{n-1}(m+n-(i+j+2))\hat{f}_{i,j}(p,q)\pi_{D,i,j} \\ \frac{1}{(m+n-2)q}\sum_{j=0}^{n-1}\sum_{i=0}^{m-1}(i+j)\hat{\nu}_{j,i}(q,p)\pi_{C,j,i}^{'} \\ \frac{1}{(m+n-2)(1-q)}\sum_{j=0}^{n-1}\sum_{i=0}^{m-1}(m+n-(i+j+2))\hat{\nu}_{j,i}(q,p)\pi_{D,j,i}^{'}\end{array}\Bigr].\]

\noindent Thus, $\overline{\pi}=p\overline{\pi}_{C}+(1-p)\overline{\pi}_{D}$ and $\overline{\pi}^{'}=q\overline{\pi}_{C}^{'}+(1-q)\overline{\pi}_{D}^{'}$ are the average fitness payoffs in the allele populations of male and female. Hence, the replicator dynamics in the group selection framework is given by the differential equations

\tiny
\begin{eqnarray*}
 \dot{p}\hspace{-1mm}&=&\hspace{-1mm}p\Bigl( \frac{1}{(m+n-2)p}\sum_{i=0}^{m-1}\sum_{j=0}^{n-1}(i+j)\hat{f}_{i,j}(p,q)\pi_{C,i,j}-\Bigl[ \frac{p}{(m+n-2)p}\sum_{i=0}^{m-1}\sum_{j=0}^{n-1}(i+j)\hat{f}_{i,j}(p,q)\pi_{C,i,j} \nonumber \\ & &+ \frac{1-p}{(m+n-2)(1-p)}\sum_{i=0}^{m-1}\sum_{j=0}^{n-1}(m+n-(i+j+2))\hat{f}_{i,j}(p,q)\pi_{D,i,j}\Bigr]  \Bigr), \nonumber \\
\dot{q}\hspace{-1mm}&=&\hspace{-1mm}q\Bigl(\frac{1}{(m+n-2)q}\sum_{j=0}^{n-1}\sum_{i=0}^{m-1}(i+j)\hat{\nu}_{j,i}(q,p)\pi_{C,j,i}^{'}-\Bigl[ \frac{q}{(m+n-2)q}\sum_{j=0}^{n-1}\sum_{i=0}^{m-1}(i+j)\hat{\nu}_{j,i}(q,p)\pi_{C,j,i}^{'} \nonumber \\ & &+\frac{1-q}{(m+n-2)(1-q)}\sum_{j=0}^{n-1}\sum_{i=0}^{m-1}(m+n-(i+j+2))\hat{\nu}_{j,i}(q,p)\pi_{D,j,i}^{'}\Bigr] \Bigr),
\end{eqnarray*}
\normalsize
on the invariant  space $[0,1]\times[0,1]$. Here, it is to be ensured  $(\hat{f}, \hat{\nu})$ satisfying the condition of Lipschitz continuous so that the system of differential equations has unique solutions through every state $(p(0),q(0))\in \Delta_{1}$ and $(q(0),p(0))\in \Delta_{2}$. For the sake of interpretation of the dynamical system, some multiple terms remain in unsimplified forms.

\section{Population structures} \label{B}
It is rare that individuals in population truly mate random. In most of the cases, populations are spatially structured and individuals are more likely to mate with other that are nearly than with  individuals from farther away (cause for geographical location). Not only that, often social structure does limit mating opportunities on others within the group. On this standpoint, in order to put mathematical tractability, it is the best choice of population structure to take a mixture of random grouping and clonal interaction. Therefore, on consideration of diploid population at autosomal locus, we get the following expressions of group frequencies: 

{\scriptsize
 \[  \Bigl[\begin{array}{ll} \hat{f}_{0,0}(p,q) & \hat{\nu}_{0,0}(q,p) \\ \hat{f}_{0,1}(p,q)& \hat{\nu}_{0,1}(q,p)\\ \hat{f}_{1,0}(p,q)& \hat{\nu}_{1,0}(q,p) \\\hat{f}_{1,1}(p,q)& \hat{\nu}_{1,1}(q,p)\end{array}\Bigr]= \Bigl[\begin{array}{ll} (1-\alpha)(1-p)(1-q)+\alpha(1-p) & (1-\alpha)(1-p)(1-q)+\alpha(1-q) \\(1-\alpha)q(1-p)&(1-\alpha)p(1-q) \\ (1-\alpha)p(1-q)& (1-\alpha)q(1-p) \\(1-\alpha)pq+\alpha p & (1-\alpha)pq+\alpha q\end{array}\Bigr]\]}

\noindent where $\alpha$ is the mating parameter. With $\alpha=0$, $\hat{f}_{i,j}$ and $\hat{\nu}_{j,i}$ adopt random selection and associated expected fitness payoffs of $C$ allele and $D$ allele in the both sexes are:

{\scriptsize
 \[ \Bigl[\begin{array}{ll} \overline{\pi}_{C} \\ \overline{\pi}_{D} \\ \overline{\pi}_{C}^{'} \\ \overline{\pi}_{D}^{'} \end{array}\Bigr]= \Bigl[\begin{array}{ll}\frac{1}{2p}(\hat{f}_{0,1}\pi_{C,0,1}+\hat{f}_{1,0}\pi_{C,1,0}+2\hat{f}_{1,1}\pi_{C,1,1}) \\ \frac{1}{2(1-p)}(2\hat{f}_{0,0}\pi_{D,0,0}+\hat{f}_{0,1}\pi_{D,0,1}+\hat{f}_{1,0}\pi_{D,1,0}) \\ \frac{1}{2q}(\hat{\nu}_{0,1}\pi_{C,0,1}^{'}+\hat{\nu}_{1,0}\pi_{C,1,0}^{'}+2\hat{\nu}_{1,1}\pi_{C,1,1}^{'})\\ \frac{1}{2(1-q)}(2\hat{\nu}_{0,0}\pi_{D,0,0}^{'}+\hat{\nu}_{0,1}\pi_{D,0,1}^{'}+\hat{\nu}_{1,0}\pi_{D,1,0}^{'})\end{array}\Bigr] = \Bigl[\begin{array}{ll} w_{11}.q+2kw_{12}.(1-q) \\ 2(1-k)w_{12}r_{pq}.q+w_{22}.(1-q)\\ v_{11}.p+2\kappa v_{21}.(1-p) \\ 2(1-\kappa)v_{21}r_{qp}.p+v_{22}.(1-p) \end{array}\Bigr]\]}
  
\noindent That is, here, the replicator dynamics of group selection does indeed encompass the replicator dynamics of game theory.
 
 At the other hand extreme point, $\alpha=1$, $\hat{f}_{i,j}$ and $\hat{\nu}_{j,i}$ follow the non-random clonal interaction framework where all selected groups are homogeneous in the sense that the groups consist of either $C$ alleles or $D$ alleles; such grouping reveals simply the clonal interaction character \citep{Bergstrom2002}.
 
 \section{ Supplemental materials } \label{C}
 Additional materials as a separate file under the heading of Supplemental Materials accompanies the paper in Elsevier Web products, including ScienceDirect: \url{http://www.sciencedirect.com}.


\end{document}